\documentclass[a4paper,fleqn,usenatbib]{mnras}
\usepackage{graphicx}
\usepackage[T1]{fontenc}
\usepackage{ae,aecompl}

\usepackage{amsmath}	% Advanced maths commands
\usepackage{amssymb}	% Extra maths symbols

\usepackage{color}

\usepackage{url}

\title[Compact Triply Eclipsing Triple TIC 209409435]{The Compact Triply Eclipsing Triple Star TIC 209409435 Discovered with \textit{TESS}}

\author[Borkovits et al.]{
T.~Borkovits$^{1,2,3}$\thanks{E-mail: borko@electra.bajaobs.hu}, 
S.~A.~Rappaport$^4$,
T.~G.~Tan$^5$,
R.~Gagliano$^6$,
T.~Jacobs$^7$,
\newauthor
X.~Huang$^{8,9}$,
T.~Mitnyan$^{1}$,
F.-J.~Hambsch$^{10}$,
T.~Kaye$^{11}$,
P.~F.~L.~Maxted$^{12}$,
\newauthor
A. P\'al$^{2}$,
A.~R.~Schmitt$^{13}$ \\
$^1$ Baja Astronomical Observatory of Szeged University, H-6500 Baja, Szegedi \'ut, Kt. 766, Hungary \\
$^2$ Konkoly Observatory, Research Centre for Astronomy and Earth Sciences, \\
 H-1121 Budapest, Konkoly Thege Miklós \'ut 15-17, Hungary \\
$^3$ ELTE Gothard Astrophysical Observatory, H-9700 Szombathely, Szent Imre h. u. 112, Hungary \\
$^4$ Department of Physics, Kavli Institute for Astrophysics and Space Research, M.I.T., Cambridge, MA 02139, USA\\
$^5$ Amateur Astronomer, Perth Exoplanet Survey Telescope, Perth, Western Australia 6010 \\
$^6$ Amateur Astronomer, Glendale, AZ 85308 \\
$^7$ Amateur Astronomer, 12812 SE 69th Place Bellevue, WA 98006 \\
$^8$ Kavli Institute for Astrophysics and Space Research, Massachusetts Institute of Technology, Cambridge, MA 02139, USA \\
$^9$ Juan Carlos Torres Fellow \\
$^{10}$ Amateur Astronomer, Remote Observatory, San Pedro de Atacama, Chile \\
$^{11}$ Amateur Astronomer, Raemor Vista Observatory, AZ \\
$^{12}$ Astrophysics Group, Keele University, Staffordshire, ST5 5BG, UK\\
$^{13}$ Citizen Scientist, 616 W. 53rd. St., Apt. 101, Minneapolis, MN 55419, USA, aschmitt@comcast.net \\ }

\begin{document}

\date{-}

\pagerange{\pageref{firstpage}--\pageref{lastpage}} \pubyear{2020}

\maketitle

\label{firstpage}

\begin{abstract}
We report the discovery in \textit{TESS} Sectors 3 and 4 of a compact triply eclipsing triple star system.  TIC 209409435 is a previously unknown eclipsing binary with a period of 5.717 days, and the presence of a third star in an outer eccentric orbit of 121.872 day period was found from two sets of third-body eclipses and from eclipse timing variations.  The latter exhibit signatures of strong 3rd-body perturbations.  After the discovery, we obtained follow-up ground-based photometric observations of several binary eclipses as well as another of the third-body eclipses.  We carried out comprehensive analyses, including the simultaneous photodynamical modelling of  \textit{TESS} and ground-based lightcurves (including both archival WASP data, and our own follow-up measurements), as well as eclipse timing variation curves.  Also, we have included in the simultaneous fits multiple star spectral energy distribution data and theoretical \texttt{PARSEC} stellar isochrones.  We find that the inner binary consists of near twin stars of mass 0.90 $M_\odot$ and radius 0.88 $R_\odot$. The third star is just 9\% more massive and 18\% larger in radius.  The inner binary has a rather small eccentricity while the outer orbit has $e = 0.40$. The inner binary and outer orbit have inclination angles within 0.1$^\circ$ and 0.2$^\circ$ of 90$^\circ$, respectively.  The mutual inclination angle is $\lesssim 1/4^\circ$.  All of these results were obtained without radial velocity observations. 
\end{abstract}  % 235 words


%%----

%----------------------------------------------------------
\section{Introduction}
\label{sec:intro}

Compact hierarchical triple-star systems are interesting because: (1) they bear on stellar formation scenarios \citep[for a concise recent review, see][Section 4, and further references therein]{czekalaetal19}; (2) long-term Kozai-Lidov cycles may drive the evolution of the inner binary \citep{lidov62,kozai62,fabryckytremaine07,toonenetal20}; (3) there can be measurable dynamical interactions that allow for masses and orbital parameters to be determined (see, e.g., \citealt{borko15}, \citealt{borko16}, \citealt{borko19a}); and (4) a complete cycle of perturbations takes place on the timescale of the outer orbit, i.e., months to years \citep{borko15}.  Clearly the latter factor is important when trying to determine system parameters on the timescale of satellite observations (months to years) for CoRoT \citep{baglin06}, \textit{Kepler} \citep{borucki10}, \textit{K2} \citep{howell14}, and \textit{TESS} \citep{ricker15}, not to mention other human timescales involving theses and careers in astronomy.  

\begin{figure*}
\centering
\includegraphics[width=0.6\textwidth]{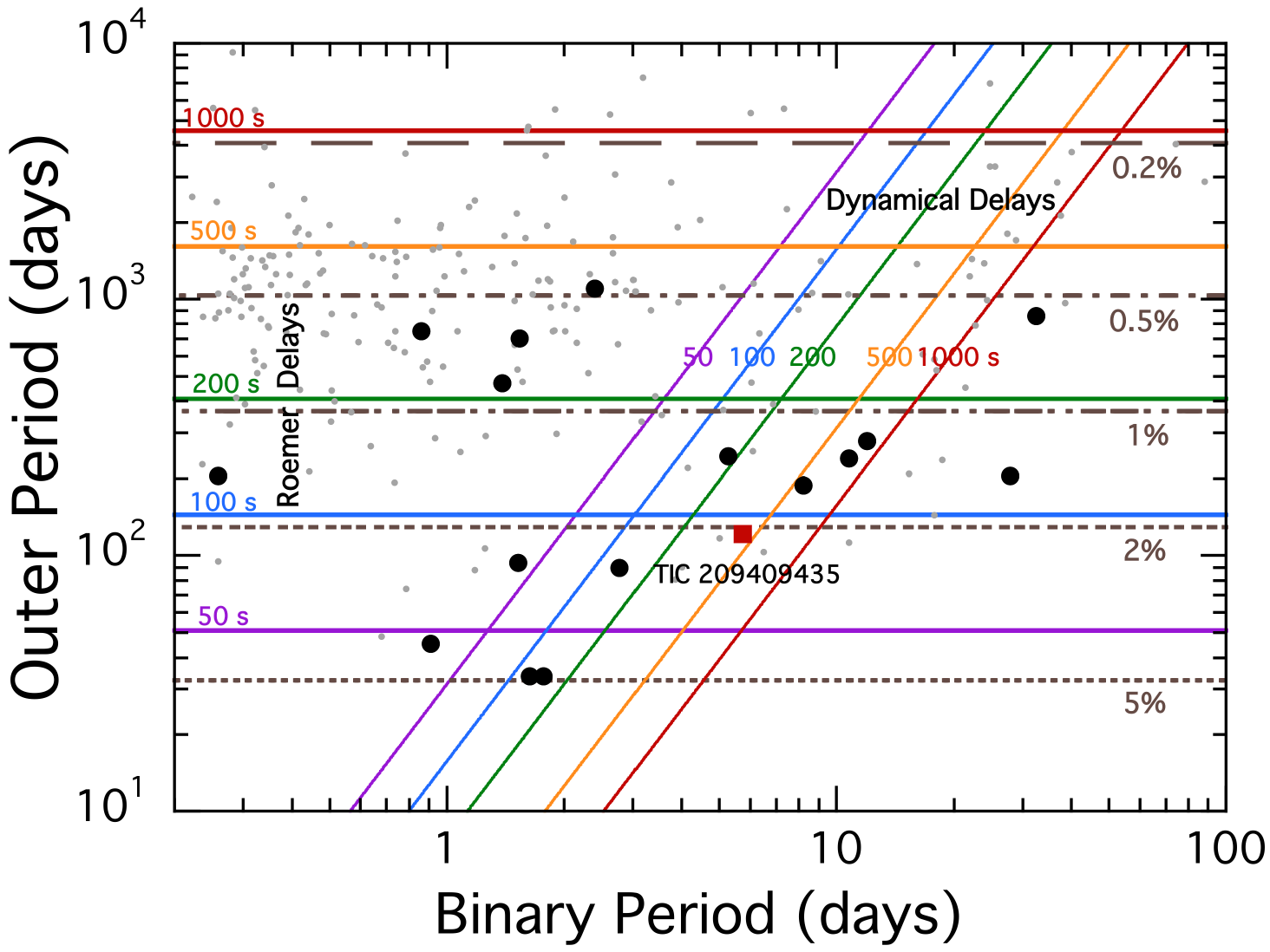}
\caption{Roemer and dynamical delays that manifest themselves in the eclipse timing variations of binary stars with third-body companions.  The results are shown in the plane of binary orbital period vs.~the period of the outer third body (i.e., of the triple star system). The colored horizontal lines are contours of typical amplitudes for the Roemer delay, while the diagonal lines are contours of typical physical delays \citep{borko15}.  The horizontal dashed lines are crude estimates of the eclipse probability for a third body orbiting the binary.  The 16 filled black circles are known triply eclipsing triple systems (see Table \ref{tbl:tripleeclipsers}), and the red square is for TIC 209409435, the system reported in this work. The fainter dots are some 200 other triple systems found with \textit{Kepler} that have measured or estimated outer orbital periods (see \citealt{borko16}).}
\label{fig:p1vsp}
\end{figure*}   % Figure 1

In this and previous work on hierarchical stellar systems we have focused on what can be learned from photometric measurements, as opposed to radial velocity studies. There are four principle photometric effects that contribute to a successful search for, and subsequent study of, these compact hierarchical triple systems: (1) binary eclipse timing variations (`ETVs') due to the classical Roemer delays (or light travel time effects); (2) ETVs due to what is referred to as ``dynamical delays", i.e., where the presence of the third body physically changes the orbital period of the inner binary; (3) the shapes of the binary eclipses; and (4) when very lucky, eclipses of the third body by the binary and vice versa.
 
The Roemer delay has an amplitude of 
\begin{equation}
A_{\rm Roem} \simeq \frac{G^{1/3}}{c(2\pi)^{2/3}}  \frac{M_\mathrm{C}}{M_{\rm ABC}^{2/3}} P_2^{2/3} (1-e_2^2 \cos\omega_2)^{1/2}\,\sin i_2
\end{equation}
(see, e.~g. \citealt{rappaport13}), where $M_C$ and $M_\mathrm{ABC}$ are the masses of the third body and the whole triple system, respectively, while $P_2$, $i_2$, $e_2$, and $\omega_2$ are the orbital period, inclination angle, eccentricity, and argument of periastron of the outer orbit of the triple star system, respectively.  By contrast, the amplitude of the dynamical delay is given approximately
\begin{equation}
A_{\rm dyn} \simeq  \frac{3}{8\pi} \frac{M_C}{M_{\rm ABC}}  \frac{P_1^2}{P_2} \, (1-e_2^2)^{-3/2} \, e_2
\end{equation}
(see \citealt{rappaport13}), for the case of near co-planarity of the orbital planes, where $P_1$ is the orbital period of the inner binary.

We illustrate the magnitudes of these delays in Fig.~\ref{fig:p1vsp}.  To keep the plot simple, we fix all the masses at 1 $M_\odot$, the eccentricity $e_2 = 0.3$, the orbit inclination angle $i_2 = 60\degr$, and $\omega_2 = 45\degr$.  The plot shows contours of constant delay amplitudes for both the Roemer and physical delays at 50, 100, 200, 500, and 1000 s.  In addition, we show contours of constant probability for outer, or third-body, eclipses.  The latter are based on assumed circular orbits and stars of 1 $R_\odot$.  

In all, with the CoRoT, \textit{Kepler}, \textit{K2}, and now \textit{TESS} missions, there are more than 200 compact triple (or quadruple) systems where ETV measurements have revealed the hierarchical nature of these systems (see, \citealt{borko16,hajduetal17} and references therein). In addition, in a very small subset of these cases, the orbital planes of the outer third star (or binary) are fortuitously aligned well enough with our line of sight so that there are so-called ``third body'' eclipsing events.  These third body eclipses greatly enhance our ability to diagnose the system parameters.  To our knowledge, there are only 17 such triply eclipsing systems\footnote{Here we do not count TIC\,167692429 for which an outer eclipse was observed in 2012, because nowdays this triple is no longer a triply eclipsing system due to the rapid precession of its orbital plane \citep[see][]{borko20}.}  with known outer periods (see Table\,\ref{tbl:tripleeclipsers}).  We show these in Fig.~\ref{fig:p1vsp} as heavy filled circles.  The outer orbits range from 34 days to 1100 days, while the binary periods cover an interval of 0.25 to 32 days. We also show in Fig.~\ref{fig:p1vsp}, as fainter points, the remaining $\sim$200 triple systems measured with \textit{Kepler} that have known or estimated outer orbital periods (see \citealt{borko16}).  

\begin{table}
\centering
\caption{List of the known close binaries exhibiting outer eclipses (in increasing order of the outer period)}
\begin{tabular}{llll}
\hline
Identifier & $P_1$ & $P_2$ & References \\
\hline
KOI-126        & 1.77 & 33.92 & 1 \\
HD~144548      & 1.63 & 33.95 & 2 \\
HD~181068      & 0.91 & 45.47 & 3 \\
CoRoT~104079133& 2.76 & 90(?) & 4 \\
KIC~4150611    & 1.52 & 94.2  & 5, 6\\
TIC~209409435  & 5.72 & 121.9 & 7 \\
EPIC~249432662 & 8.19 & 188.4 & 8 \\
KIC~2856960    & 0.26 & 204.8 & 9, 10\\
KIC~7668648    & 27.83& 204.8 & 11, 12\\
KIC~6964043    & 10.73& 239.1 & 11\\
KIC~7289157    & 5.27 & 243.4 & 11, 12\\
OGLE-BLG-ECL-187370&11.96&280.5& 13\\
KIC~9007918    & 1.39 & 470.9 & 14 \\
b~Persei       & 1.52 & 704.5 & 15\\
KIC~2835289    & 0.86 & 755   & 16\\
KIC~5255552    & 32.47& 862.1 & 11\\ 
KIC~6543674    & 2.39 & 1101.4& 11, 17\\
\hline
\end{tabular}

{\bf References:} (1) \citet{carteretal11}; (2) \citet{alonsoetal15}; (3) \citet{derekasetal11}; (4) \citet{hajduetal17}; (5) \citet{TheThing}; (6) \citet{helminiaketal17}; (7) This paper; (8) \citet{borko19a}; (9) \citet{armstrongetal12}; (10) \citet{marshetal14}; (11) \citet{borko15}; (12) \citet{orosz15}; (13) unpublished, ongoing analysis; (14) \citet{borko16}; (15) \citet{collinsetal14} (16) \citet{conroyetal14}; (17) \citet{masudaetal15}
\label{tbl:tripleeclipsers}
\end{table}

In the triply eclipsing systems, when eclipse timing variations are combined with a photodynamical analysis (see, e.g., \citealt{borko19a}), including spectral energy distributions (`SEDs'), stellar isochrone models, and the Gaia distance, many of the system parameters can be determined.  This includes all three masses, periods, eccentricities, and inclination angles of the various orbital planes. 

\begin{table}
\centering
\caption{Archival Properties of the TIC 209409435 Triple System}
\begin{tabular}{lc}
\hline
\hline
Parameter & Value   \\
\hline
RA (J2000) & 45.21625  \\  
Dec (J2000) &  $-34.45707$  \\  
$T$$^a$ & $13.208 \pm 0.006$ \\
$G$$^c$& $13.647 \pm 0.002$  \\
$G_{\rm BP}$$^c$ & $13.985 \pm 0.001$  \\
$G_{\rm RP}$$^c$ & $13.149 \pm 0.001$  \\
B$^a$ & $14.427 \pm 0.019$ \\
V$^b$ & $13.777 \pm 0.018$ \\
g$'^b$& $14.065 \pm 0.018$ \\
r$'^b$& $13.613 \pm 0.029$ \\
i$'^b$& $13.435 \pm 0.047$ \\
J$^d$ & $12.622 \pm 0.024$   \\
H$^d$ & $12.267 \pm 0.024$  \\
K$^d$ & $12.222 \pm 0.023$  \\
W1$^e$ & $12.155 \pm 0.023$ \\
W2$^e$ & $12.193 \pm 0.022$  \\
W3$^e$ & $11.973 \pm 0.19$  \\
W4$^e$ & $> 9.464$  \\
$T_{\rm eff}$ (K)$^c$ & $5795 \pm 30$  \\
%$R$ ($R_\odot$)$^b$ & $1.89 \pm 0.2 $  \\
%$M$ ($M_\odot$)$^e$ & $2.4 \pm 0.2$ \\
$L$ ($L_\odot$)$^c$ & $2.47 \pm 0.1 $  \\
Distance (pc)$^f$ & $949\pm15$  \\   
$\mu_\alpha$ (mas ~${\rm yr}^{-1}$)$^c$ & $-1.09 \pm 0.02$   \\ 
$\mu_\delta$ (mas ~${\rm yr}^{-1}$)$^c$ &  $+2.77 \pm 0.03$   \\ 
\hline
\label{tbl:mags}  % Table 1
\end{tabular}

\textit{Notes.}  (a) TIC-8 catalog \citep{TIC}. (b) AAVSO Photometric All Sky Survey (APASS) DR9, \citep{APASS}, \url{http://vizier.u-strasbg.fr/viz-bin/VizieR?-source=II/336/apass9}. (c) Gaia DR2 \citep{GaiaDR2}.  (d) 2MASS catalog \citep{2MASS}.  (e) WISE point source catalog \citep{WISE}. (f) \citet{bailer-jonesetal18}.
\end{table}

\begin{figure*}
\begin{center}
\includegraphics[width=0.75 \textwidth]{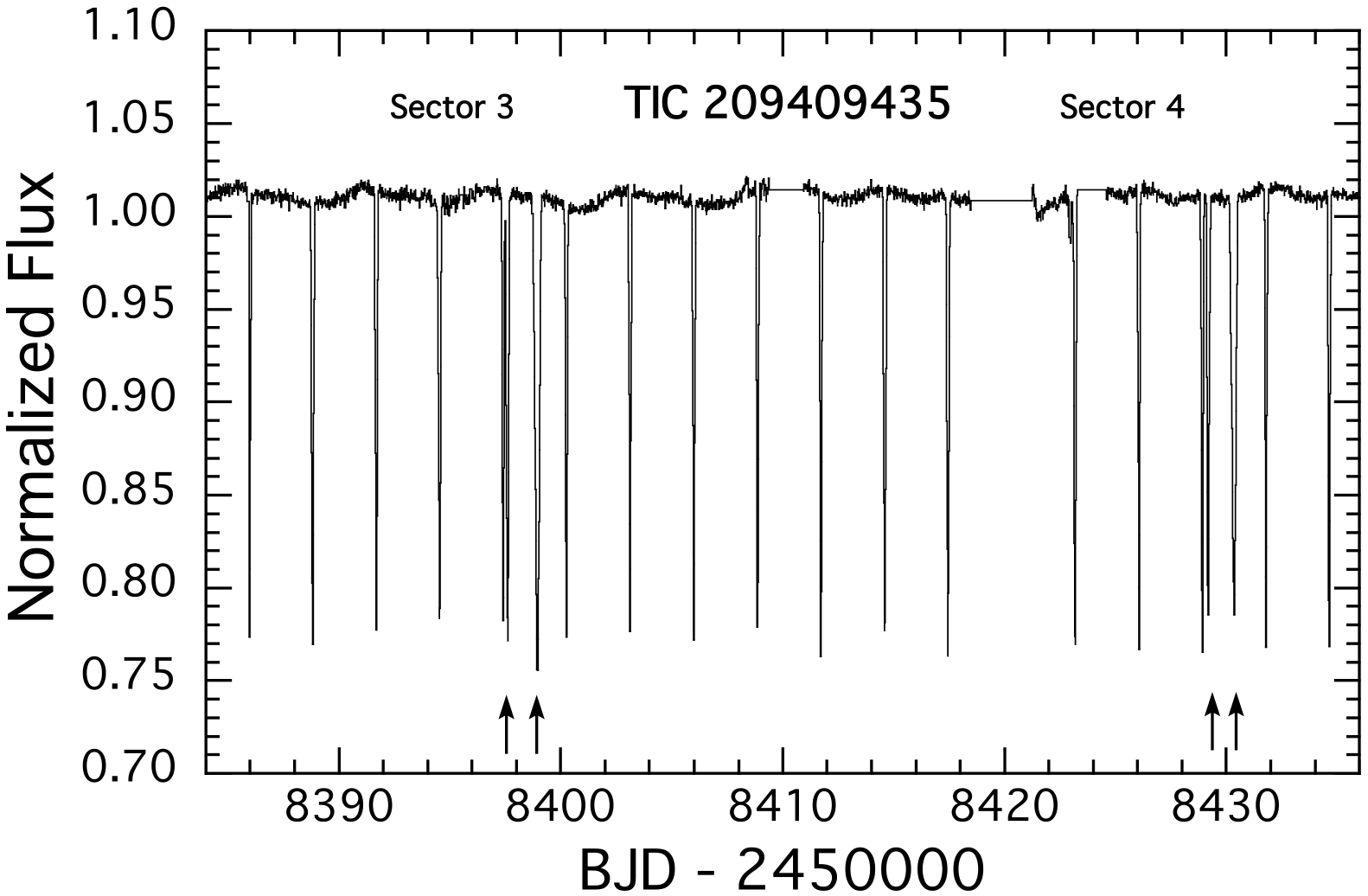}
\caption{The \textit{TESS} lightcurve from Sector 3, in which it was first discovered, and Sector 4.  The full frame images have 30-minute cadence.  The four arrows mark the times of the anomalous third-body eclipses.  The first two of these correspond to the inner binary eclipsing the third star, while the second set is the opposite situation where the 3rd star eclipses the inner binary. }
\label{fig:rawLC} 
\end{center}
\end{figure*} % Figure 2

In this paper, we report on TIC 209409435, the first of the compact triply eclipsing triples found with \textit{TESS}, plus a full orbital solution for the system.  The system consists of a 5.7-day binary in orbit with a third star in a 122-day outer orbit. The  photometric properties of the composite system are summarized in Table \ref{tbl:mags}. In Section 2 we describes all the available observational data, as well as their preparation for the analysis. Then, Section 3 provides a full explanation of the steps of the joint physical and dynamical modeling of the light- and ETV curves, SED, parallax and stellar isochrones. In Section  4 we discuss the results from astrophysical and dynamical points of views. Finally, in Sect. 5 we summarize our findings and draw conclusions from our work.

%%--------------------------------------------------------------

\section{Observational data}
\label{sec:obs}

\subsection{\textit{TESS} Observations of TIC 209409435}
\label{sec:observ}

\begin{figure}
\begin{center}
\includegraphics[width=0.49 \textwidth]{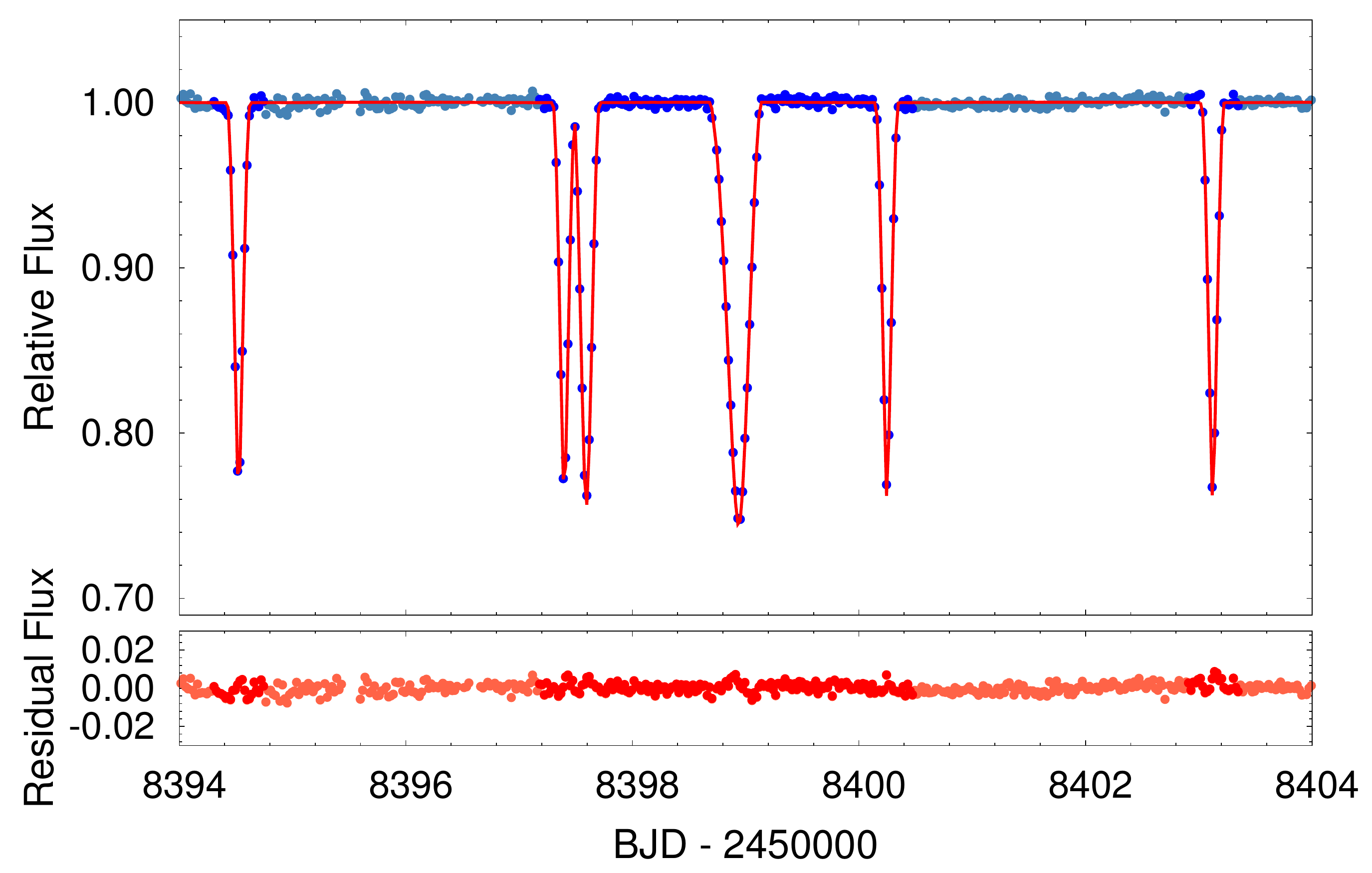}
\includegraphics[width=0.49 \textwidth]{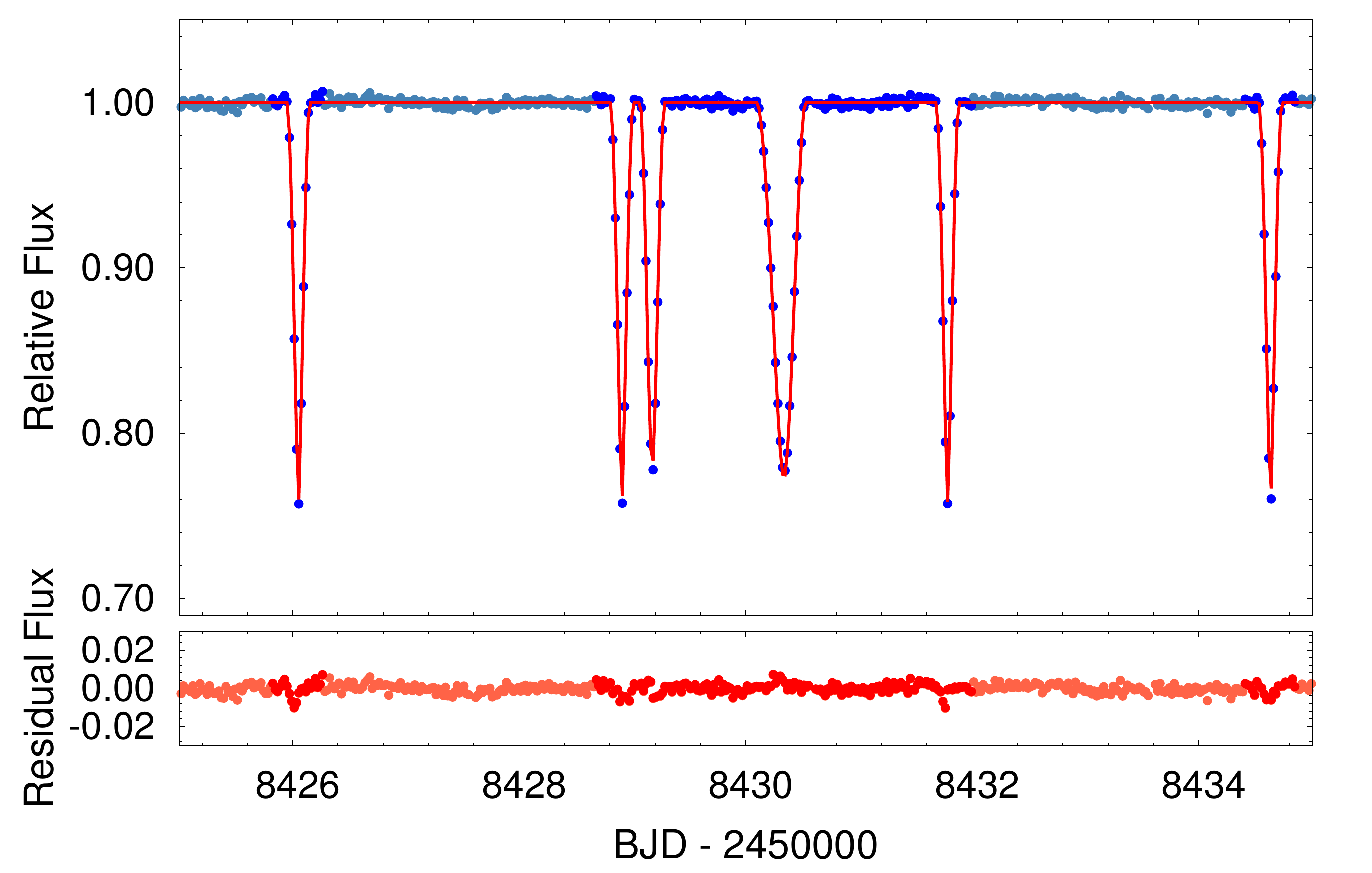}
\caption{Expanded view of the two sets of anomalous third-body eclipses. The dark blue circles represent those data points that were used for the photodynamical model (see Sect.~\ref{sec:dyn}), while the other out-of-eclipse data (not used in the modelling) are plotted as pale blue circles. The red curve is the photodynamical model solution corrected for the 30-minute integration time; the residuals to the model are also shown below the lightcurves. The top panel shows the first two events where the inner binary passes in front of the third star and produces two eclipses in addition to the regular binary eclipses spaced every $5.717\div 2$ days.  The broader of the anomalous eclipses occurs in between two regular eclipses where the transverse motion of the binary stars across the sky is at a relative minimum.  The bottom panel shows the two anomalous eclipses when the third star passes in front of the inner binary.  By chance circumstances, the pattern of the two anomalous events in the top and bottom panels, with respect to the regular eclipses, is nearly identical.} 
\label{fig:outecl} 
\end{center}
\end{figure}  % Figure 3

\begin{figure}
\begin{center}
\includegraphics[width=0.99\columnwidth]{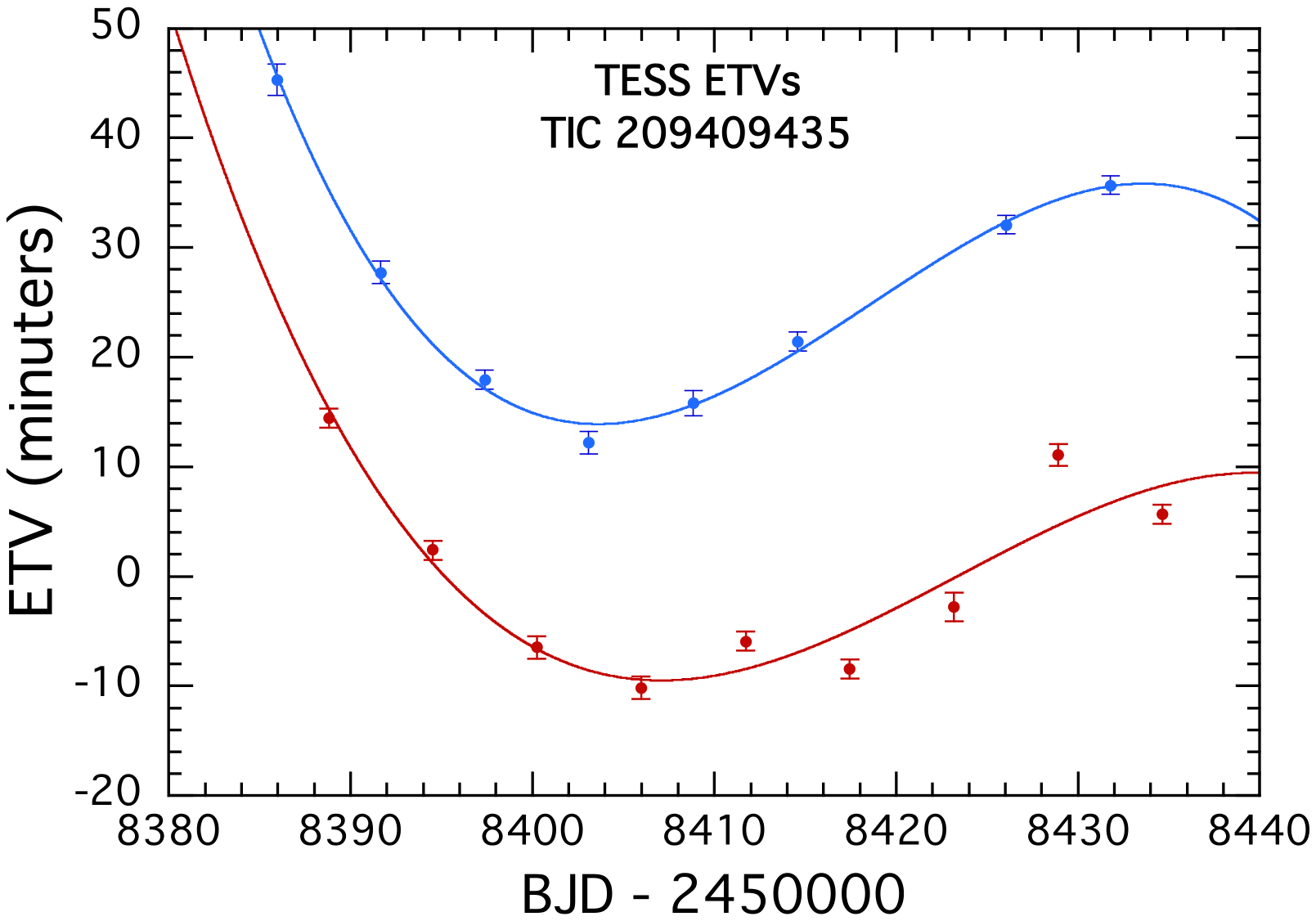}
\caption{ Eclipse timing variations of TIC 209409435 during the \textit{TESS} discovery observations.  The red and blue points are for the primary and secondary eclipses, respectively.  The highly significant non-linear behavior is a clear indication that the third star is perturbing the binary.  The smooth curves are fits to an ad hoc third-order polynomial just to guide the eye.}
\label{fig:ETV}  
\end{center}  
\end{figure}  % Figure 4

\begin{figure}
\begin{center}
\includegraphics[width=0.49\textwidth]{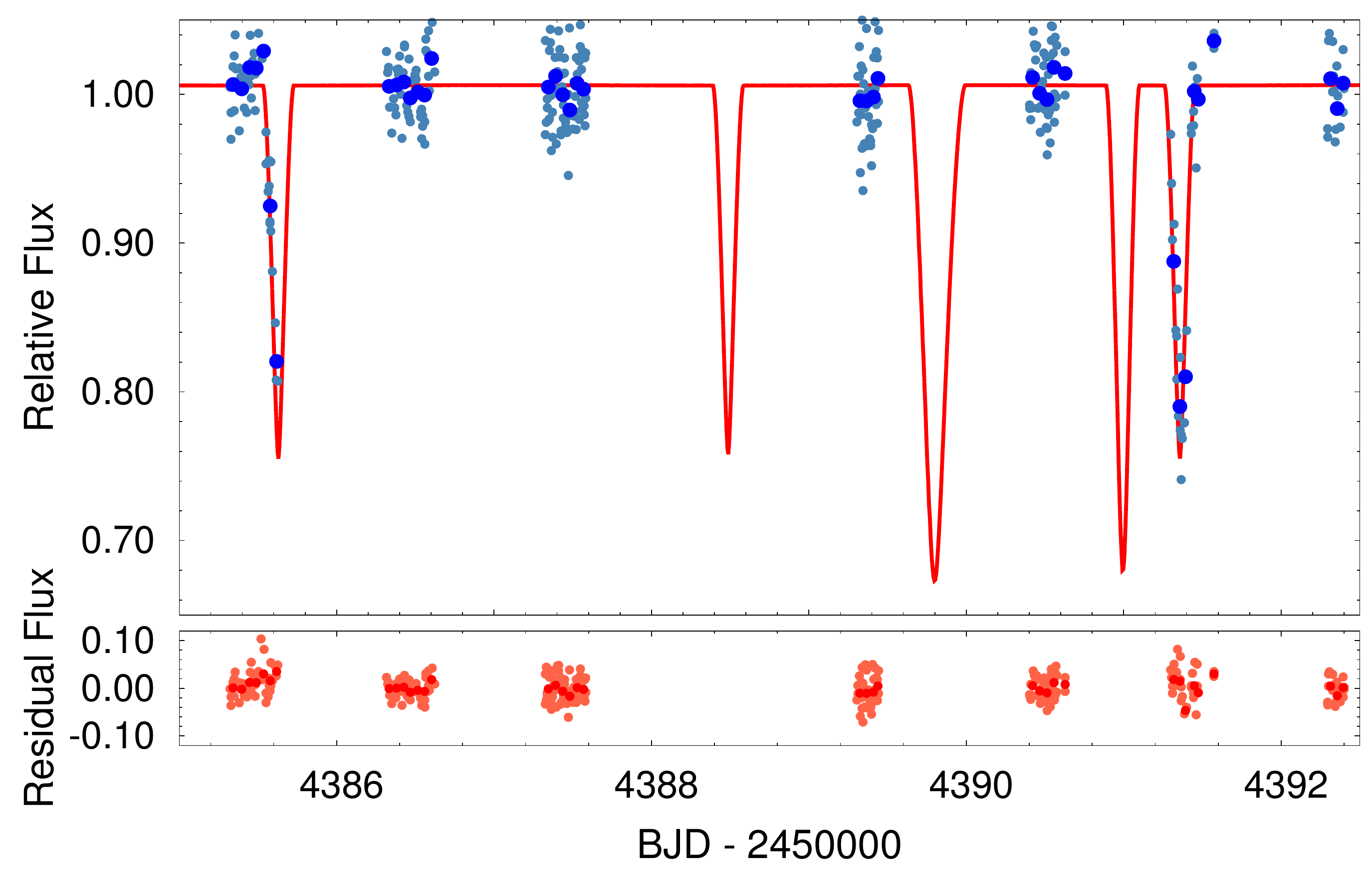}
 \includegraphics[width=0.49\textwidth]{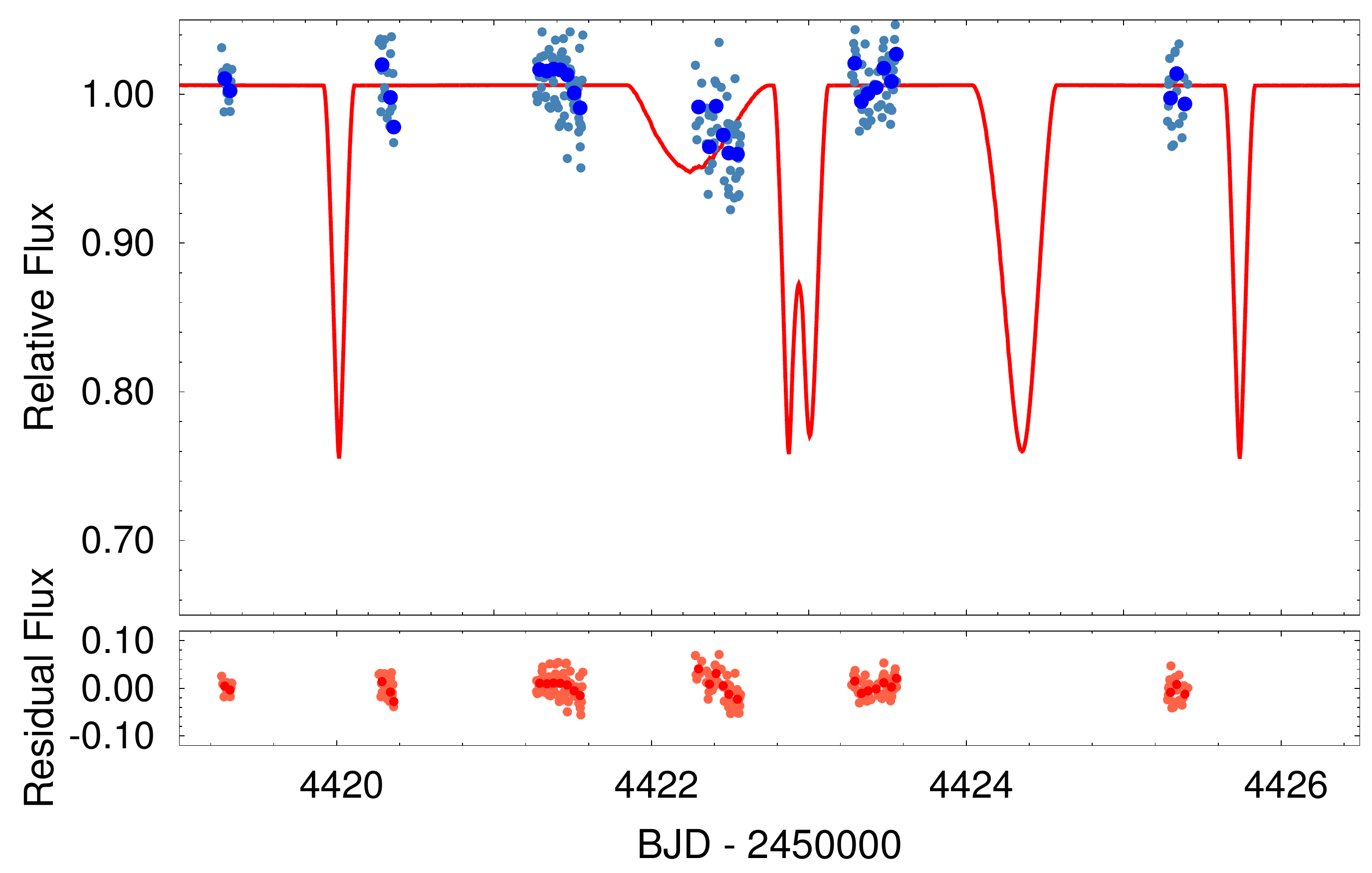}
\caption{Two sections of SuperWASP data for TIC\,209209435. The pale and dark blue points are the original and the 1-hr averaged WASP photometric data for this target, respectively.  Only the dark blue points were used in the fit.  The solid red curve is the photodynamical model back-projected more than a decade from the recent \textit{TESS} observations.}
\label{fig:wasp_lc} 
\end{center}
\end{figure} % Figure 5

\begin{figure*}
\begin{center}
\includegraphics[width=0.60\textwidth]{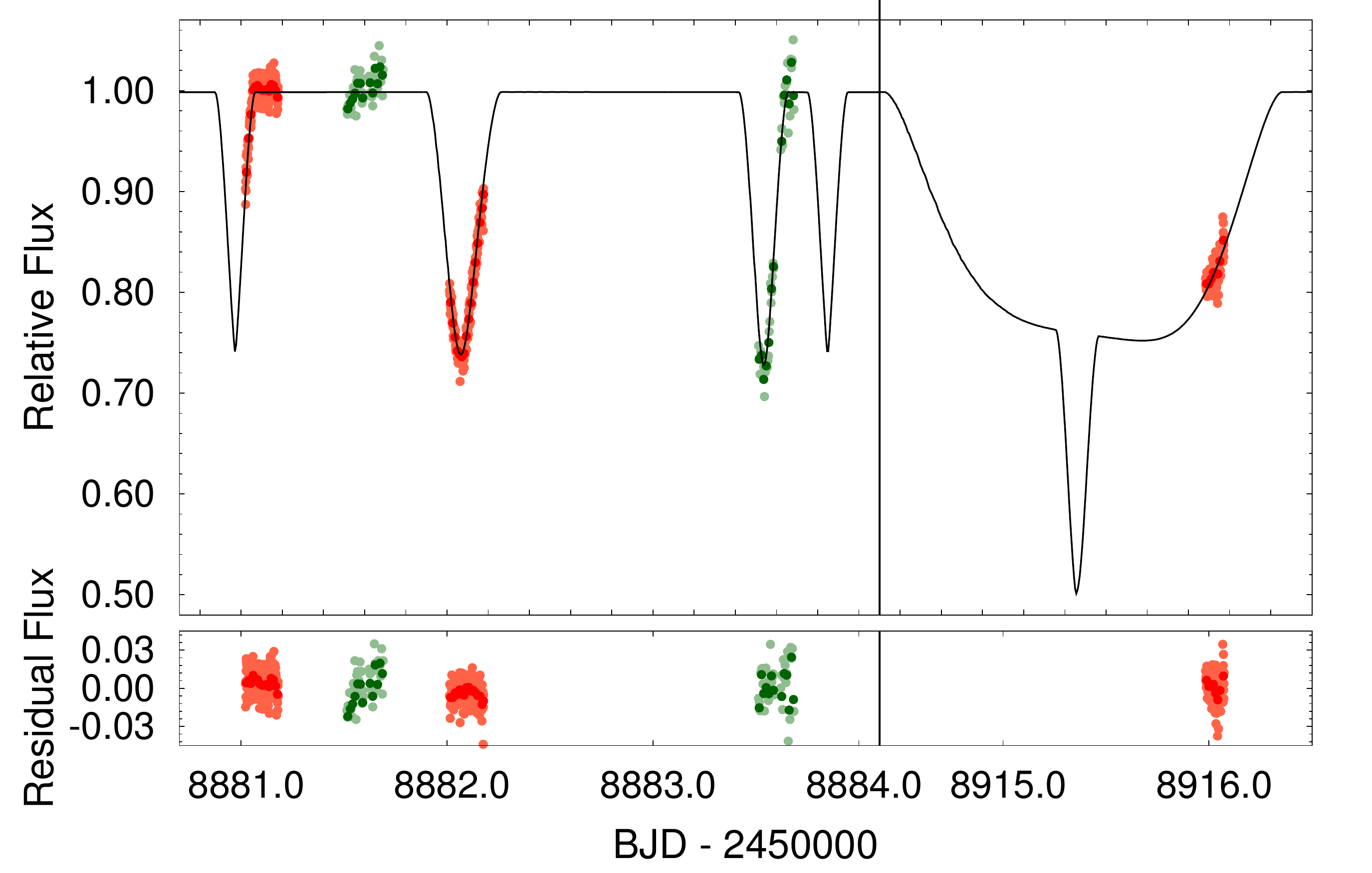}
\caption{Ground based photometric follow-up observations of TIC\,209409435 right before and during its February and March 2020 outer eclipsing events.  Red and green data points represent the observations in Cousins $R_\mathrm{C}$- (TG Tan at Perth) and Johnson $V$-bands (F.-J. Hambsch at ROAD), respectively. (Dots with lighter colors display the original, observed data points, while the darker ones represent the binned points, used for the analysis, see, in Sect.\,\ref{sec:dyn}.)  The black line is the photodynamical model light curve in $R_\mathrm{C}$-band (see Sect.\,\ref{sec:dyn}). The early February event was similar to those that were observed by \textit{TESS}, i.~e., two separate outer eclipses (around times 8882.0 and 8883.5) which occurred between two regular eclipses, indicating that the two members of the inner binary eclipsed the outer component separately.  By contrast the next outer eclipse event, between 8914 and 8916 displayed a long duration, large amplitude, irregular shaped dip, resembling the one that was observed by \textit{Kepler} in the triple system EPIC\,249432662 \citep{borko19a}. In this event the outer star eclipsed the binary members during their conjunction and, therefore, due to the almost coplanar configuration, the secondary of the inner binary had a small velocity relative to the outer star, resulting in the long-lasting eclipse, while the inner pair's primary, moving into the opposite direction, produced the sharp dip at the bottom of the main event.}
\label{fig:lc_ground} 
\end{center}
\end{figure*}  % Figure 6

In addition to conventional computer searches through the \textit{TESS} data for periodic signals, e.g., due to transiting planets and eclipsing binaries, a group of us has been visually surveying all the the full-frame image (`FFI') stars down to about \textit{TESS} magnitude 13.5.  The lightcurves that we use for surveying the data are from the MIT Quicklook pipeline (`QLP';  \citealt{huang19}).  For each star, five lightcurves are extracted from apertures with different sizes ranging from $1.75 - 8$ pixels\footnote{Each \textit{TESS} pixel is $20'' \times 20''$}.  The best aperture is chosen for stars in a particular magnitude bin based on the photometric precision after detrending. Fainter stars have relatively smaller photometric apertures. The 1.75-pixel apertures are almost never used because our method is difference-image-based, and we automatically deblend the source flux based on the target's T magnitude. This requires that we have an aperture that includes most of the light from the source, and the 1.75-pixel aperture is not really sufficient for that.

TIC\,209409435 was observed by the \textit{TESS} spacecraft \citep{ricker15} during Year 1 in  Sectors 3 and 4.  No 2 minute cadence data were available, thus, this triple-star system was observed serendipitously as part of the FFI coverage of those sectors. For the FFIs, the observational cadence is 30 minutes.

While surveying these images using the {\tt LcTools} software system \citep{schmitt19}, two of us (RG and TJ) independently discovered the triple nature of this source (on 19 and 24 November, 2019).  The QLP \textit{TESS} lightcurve is shown in Fig.~\ref{fig:rawLC}.  Aside from an unremarkable 5.7-day binary with two nearly equal eclipses per orbit, it is apparent that there are four anomalous eclipses (marked with arrows in the figure). The two sets of closely spaced anomalous eclipses are separated by $\sim$32 days.  The anomalous eclipses are shown in more detail in Fig.~\ref{fig:outecl}.  The red curve is a model fit which will be discussed later in the paper.  Because of the temporal pattern of the anomalous eclipses, we immediately suspected that these were due to a third body orbiting the binary.

We then computed an eclipse timing variations (`ETV') curve for the 17 available eclipses (see Fig.~\ref{fig:rawLC}).  The results are shown in Fig.~\ref{fig:ETV}, and tabulated in Table\,\ref{Tab:TIC_209409435_ToM}. (Note, due to a lack of data points on the ingress of  the very first eclipse, it was dropped from the ETV curve used for the complex photodynamical analysis in Sect.\,\ref{sec:dyn}, and not listed in the table.)  The red and blue points represent the primary and secondary eclipses, respectively, while the smooth curves are fits to a cubic function just to guide the eye.  The offset between the two curves by $\sim$20 minutes indicates that the binary orbit has a small eccentricity.  The non-linear behavior of the ETVs, over just the 50-day interval of the \textit{TESS} observations, confirms that there are dynamical interactions with the third body.  

\subsection{WASP Observations}

Fortuitously for our study, TIC\,209409435 was observed within the field of the WASP-South project \citep{2006PASP..118.1407P,2006MNRAS.373..799C} during three seasons between June 2006 and January 2012. (There was a huge gap in the observations between January 2008 and August 2011.)  The WASP instruments each consist of an array of 8 cameras with Canon 200-mm f/1.8 lenses and  2k$\times$2k $e$2$V$ CCD detectors providing images with a field-of-view  of $7.8^{\circ}\times 7.8^{\circ}$ at an image scale of 13.7 arcsec/pixel. Images are obtained through a broad-band filter covering 400-700\,nm. Fluxes are measured in an aperture with a radius of 48 arcsec for the WASP data. The data are processed with the SYSRem algorithm \citep{2005MNRAS.356.1466T} to remove instrumental effects. 

Almost two dozen eclipses can be identified in the WASP archival observations of TIC\,209409435.  Although many of these eclipses were observed only partially and, therefore, we were able to determine mid-eclipse times for only a portion of them (see in Table\,\ref{Tab:TIC_209409435_ToM}), these data were essential for our analysis. In particular, they helped to constrain: (i) the outer orbital period -- not only through the ETV of the regular eclipses, but also by narrowing the possible locations of the outer eclipses;  (ii) the `flatness' of the triple system through the presence, or the lack of eclipse depth variations which might be forced by nodal precession in case of non-coplanarity of the inner and outer orbital planes; and also (iii) the dynamically forced apsidal motion, and in such a way, the dynamical properties of this strongly gravitationally interacting triple. Therefore, we used the entire WASP lightcurve for the complex photodynamical analysis (see Sect.\,\ref{sec:dyn}). We plot two sections of the WASP data in Fig.\,\ref{fig:wasp_lc}.

\subsection{Follow-up Ground-Based Observations}

Following the discovery of this triply eclipsing triple star system we organized a photometric follow-up observational campaign with the participation of three amateur astronomers operating their own private observatories.  We describe the observatories and the observations in the following subsections.

\subsubsection{Ground-Based Observatories}

\textit{PEST Observatory:} The observatory is owned and operated by Thiam-Guan (TG) Tan. PEST is equipped with a 12-inch Meade LX200 SCT f/10 telescope, an SBIG ST-8XME CCD camera, a BVRI filter wheel, a focal reducer yielding f/5, and an Optec TCF-Si focuser controlled by the observatory computer. PEST has a $31' \times 21'$ field of view and a $1.2''$ per pixel scale. PEST is located in a suburb of the city of Perth, Western Australia. Observations of TIC\,209409435 were carried out in Cousins $R_\mathrm{C}$-band, with an exposure time of 60 sec.  A custom pipeline based on {\tt C-Munipack} was used to calibrate the images and extract the differential time-series photometry using a photometric aperture of 6.2$''$ radius.
 
\textit{Raemor Vista Observatory and Junk Bond Observatory:} The observations were conducted by T.G.~Kaye and the image sets were calibrated and measured by B.~Gary and T.G.~Kaye.  Raemor Vista has a 0.4-m Dreamscope with an Apogee CG 16 M CCD camera.  30-s exposures were used, and the images were spectrally unfiltered and binned into $2 \times 2$ pixels.  The Junk Bond Observatory has a 0.8-m Ritchey Chr\'etien telescope with an SBIG STL6303E CCD camera.  30-s exposures were used, and the images were also unfiltered and binned into $2 \times 2$ pixels.

\textit{ROAD Observatory} (Remote Observatory Atacama Desert): The observatory is located in San Pedro de Atacama, Chile and operated by F.-J. Hambsch remotely from Belgium. It uses a 16-inch telescope with a FLI ML16803 CCD camera having a $4096 \times 4096$ Kodak KAF-16803 image sensor. The exposure were 60 s with a Johnson V filter. Aperture photometry using the freely available software {\tt LESVEPHOTOMETRY} was use together with reference and check stars. Dark and sky flat calibration frames were applied to the images.

\subsubsection{Measured Ground-Based Eclipses}

Observations were taken on 13 nights between 30 November 2019 and 7 March 2020. During the first six weeks of this campaign we were able to observe three regular eclipses of the inner binary (see Table\,\ref{Tab:TIC_209409435_ToM}), and an almost complete ingress portion of an additional secondary inner eclipse. These observations were essential for the preliminary photodynamical modelling (see Sect.\,\ref{sec:dyn}) which made it possible to constrain the outer period of the triple and to predict the occurrence time of the forthcoming extra (third-body) eclipses with an accuracy of a few hours. Thanks to these predictions we were able to successfully observe and detect the upcoming third-body eclipse events which occurred relatively close to their predicted times (see Fig.\,\ref{fig:lc_ground}). %Note,  even the chance factor gave us a big help during the first predicted extra eclipse at 1 February 2020, as it was delayed by 6 hours to the model prediction which resulted in a much better coverage for our observations.  
As it turned out, the first predicted third-body eclipse on 1 February 2020 was late in arriving by 6 hours, and this fortuitously gave us the longitude coverage on the Earth to detect it.

\begin{figure}
\begin{center}
\includegraphics[width=1.00\columnwidth]{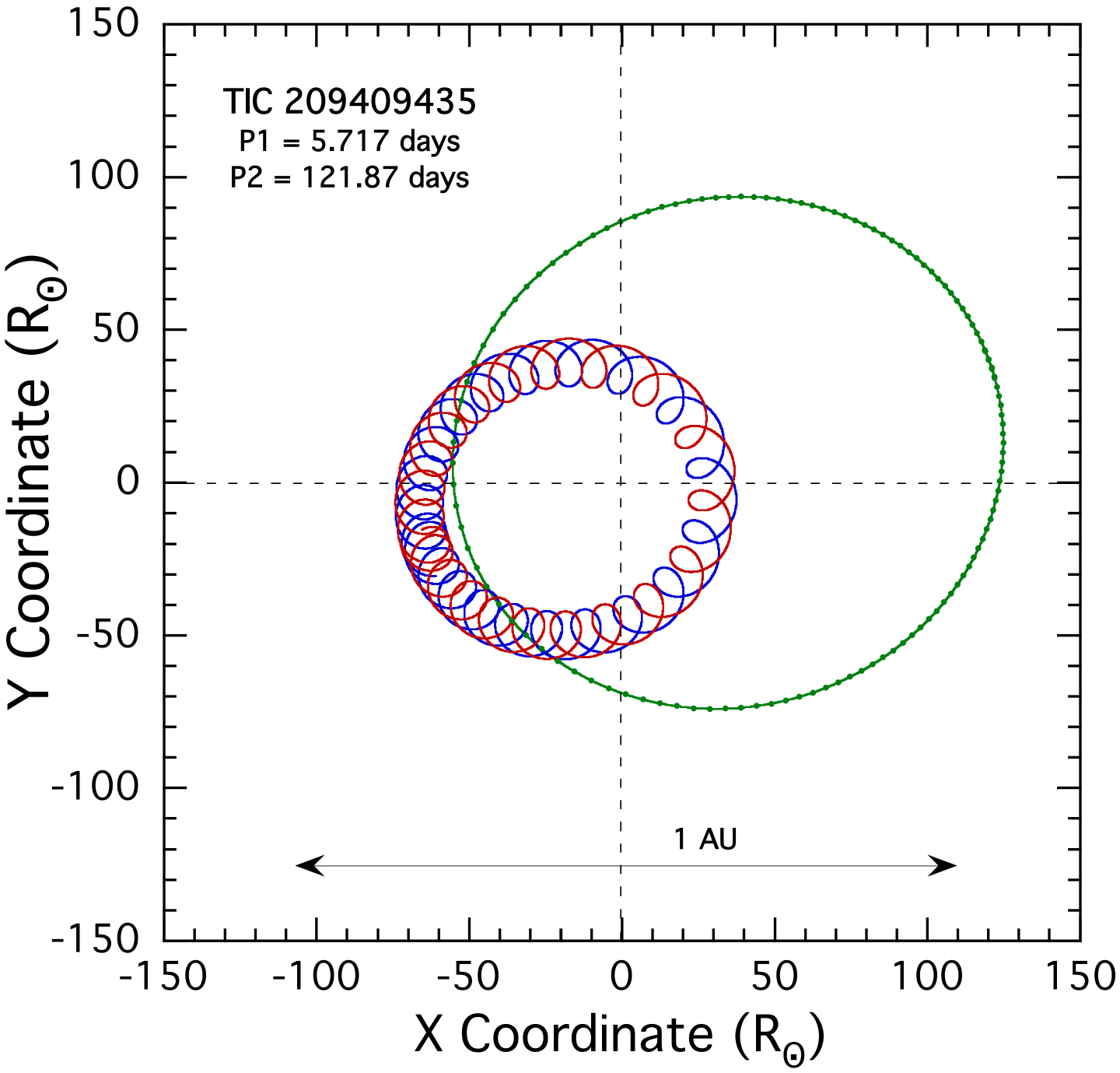}
\caption{To-scale drawing of the orbits of the three stars in the TIC 209409435 triple system. Red and blue lines represent the inner binary with its 5.717-d period, while the green curve is the orbit of the outer third body with a period of 121.87 days.  The system center of mass is at \{0,0\} and the observer is viewing the system from a distant position along the $-Y$ axis.  All orbits are moving counterclockwise. The green dots along the outer orbit are spaced by 1 day. }
\label{fig:orbit}
\end{center}
\end{figure}  % Figure 7

\begin{table*}
\caption{Times of minima of TIC 209409435}
 \label{Tab:TIC_209409435_ToM}
\begin{tabular}{@{}lrllrllrl}
\hline
BJD & Cycle  & std. dev. & BJD & Cycle  & std. dev. & BJD & Cycle  & std. dev. \\ 
$-2\,400\,000$ & no. &   \multicolumn{1}{c}{$(d)$} & $-2\,400\,000$ & no. &   \multicolumn{1}{c}{$(d)$} & $-2\,400\,000$ & no. &   \multicolumn{1}{c}{$(d)$} \\ 
\hline
53953.574400 & -775.0 & 0.003295 & 55899.366840 & -435.0 & 0.015246 & 58414.588636 &    4.5 & 0.001370 \\ 
53993.623179 & -768.0 & 0.010601 & 58388.803073 &    0.0 & 0.000565 & 58417.434057 &    5.0 & 0.000672 \\ 
53996.487379 & -767.5 & 0.001758 & 58391.677119 &    0.5 & 0.000589 & 58423.167677 &    6.0 & 0.002938 \\ 
54036.557633 & -760.5 & 0.025052 & 58394.524148 &    1.0 & 0.001387 & 58426.053529 &    6.5 & 0.000701 \\ 
54365.608594 & -703.0 & 0.002298 & 58397.398966 &    1.5 & 0.000756 & 58428.907548 &    7.0 & 0.000853 \\ 
54368.487180 & -702.5 & 0.023878 & 58400.247192 &    2.0 & 0.000516 & 58431.785006 &    7.5 & 0.014449 \\ 
54391.368020 & -698.5 & 0.002840 & 58403.123639 &    2.5 & 0.006389 & 58434.632628 &    8.0 & 0.000910 \\ 
54411.414138 & -695.0 & 0.002780 & 58405.973836 &    3.0 & 0.000455 & 58818.060304 &   75.0 & 0.000156 \\ 
54414.281903 & -694.5 & 0.004802 & 58408.855770 &    3.5 & 0.000880 & 58838.092983 &   78.5 & 0.000154 \\ 
54454.330712 & -687.5 & 0.004301 & 58411.706419 &    4.0 & 0.000672 & 58858.091655 &   82.0 & 0.000239 \\ 
\hline
\end{tabular}

\textit{Notes.} Integer and half-integer cycle numbers refer to primary and secondary eclipses, respectively. Eclipses between cycle nos. $-755.0$ and $-435.0$ were observed in the WASP project. Eclipse times between cycle nos. $0.0$ and $8.0$ were determined from the \textit{TESS} measurements. Finally, the last three eclipses (cycle nos. $75.0-82.0$) were observed at PEST observatory within the timeframe of our  photometric follow-up campaign.
\end{table*}

\subsection{Overview of the Triple System}

Based on the lightcurve information discussed in the previous subsections, we have carried out a detailed photodynamical evaluation of the system parameters.  This is described in detail in Section \ref{sec:dyn}, which follows next. For now we simply present a to-scale drawing of what the TIC\,209409435 triple-star system looks like---see Fig.\ref{fig:orbit}.  The red and blue lines in Fig.\,\ref{fig:orbit} represent the inner binary with its 5.717-d period, while the green curve is the orbit of the outer third body with a period of 121.87 days.  The observer is viewing the system from a distant position along the $-Y$ axis.  The outer orbit is sufficiently eccentric that, at least according to the drawing, it is easy to understand why the intervals between 3rd body events are not nearly equally spaced.

We next describe how the photometry data alone, without any radial velocity measurements, are sufficient to deduce the constituent stellar masses and system geometry with remarkable fidelity.

\begin{figure*}
\begin{center}
\includegraphics[width=0.99 \textwidth]{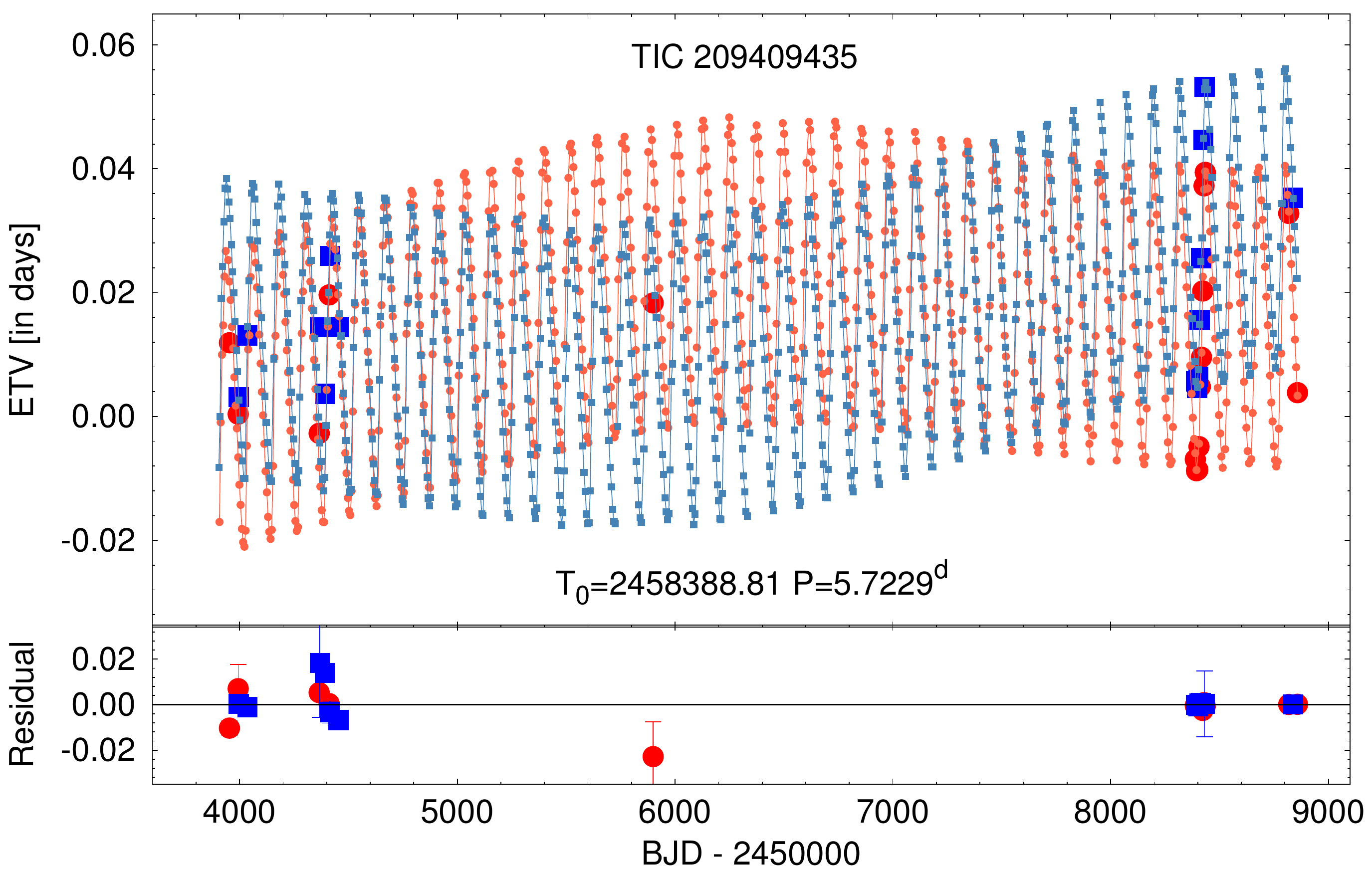}
\includegraphics[width=0.47 \textwidth]{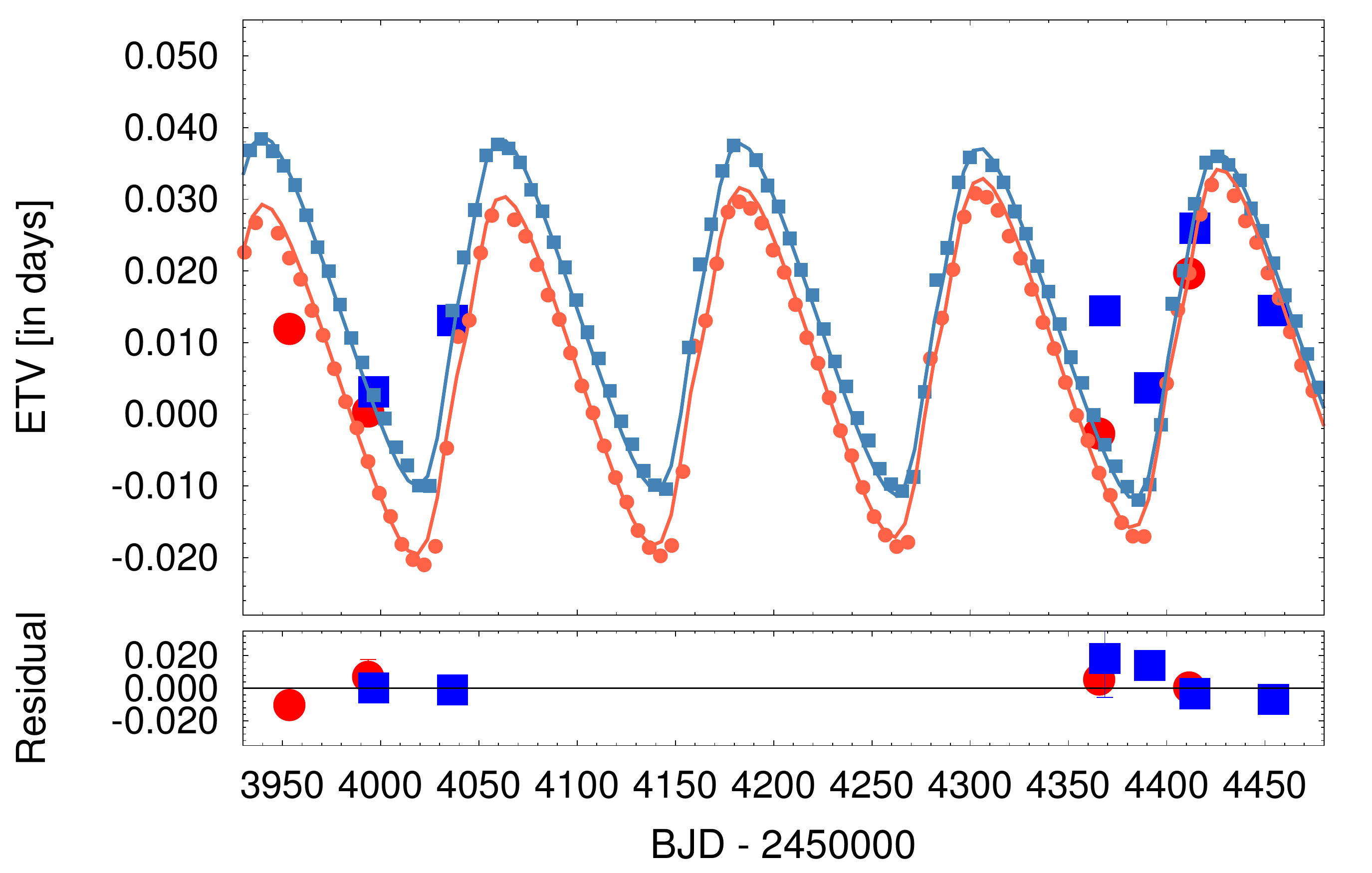}\includegraphics[width=0.47 \textwidth]{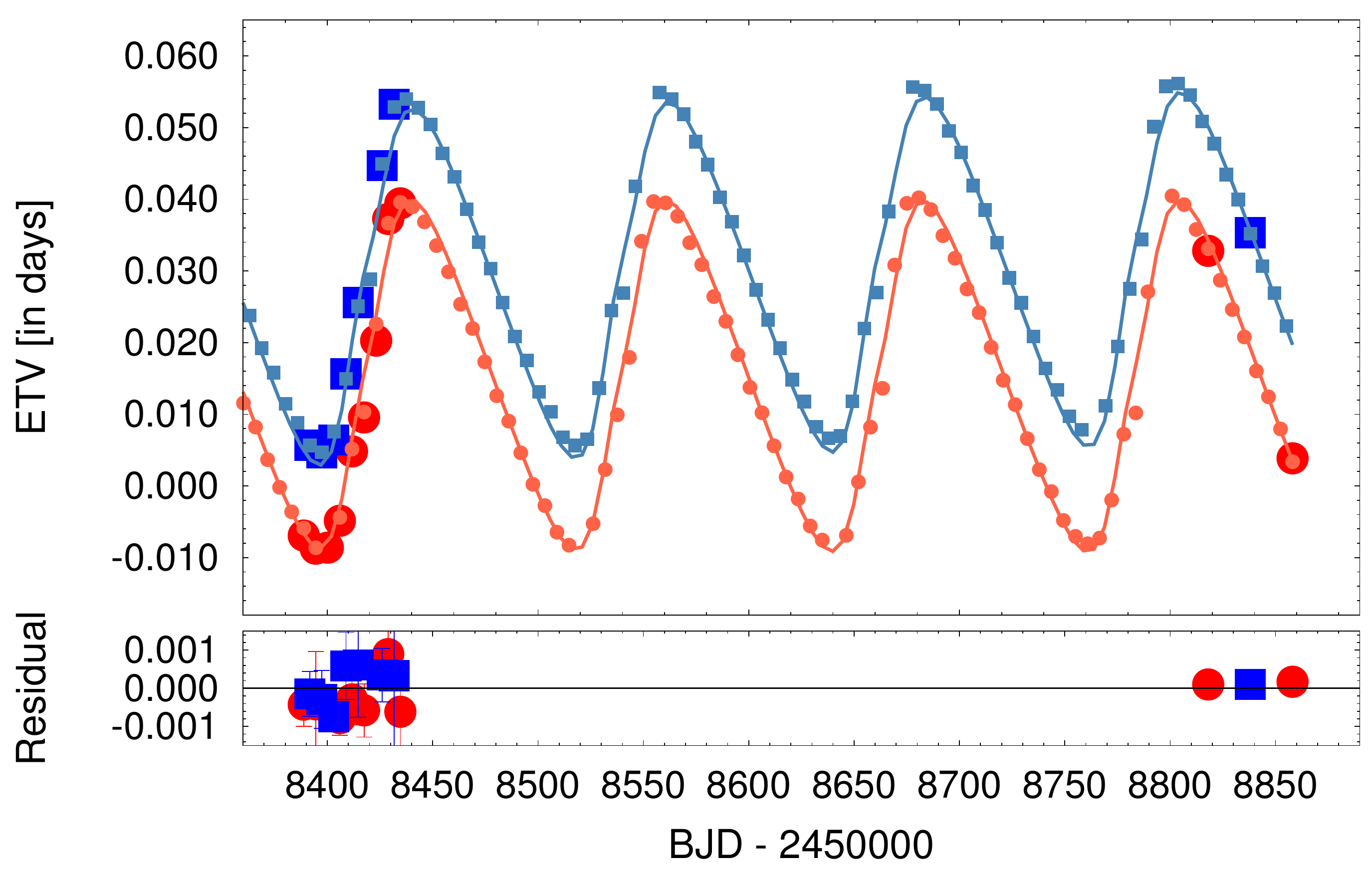}
\caption{Eclipse timing variations of TIC\,209409435. The large red filled circles and blue squares are calculated from the observed eclipse events, while the corresponding smaller symbols with lighter colors are determined from the photodynamical model solution. These model ETV points are connected to each other simply to guide the reader's eye.  For better visibility, bottom panels show zoom-ins of the ETV curves for the epochs of the SuperWASP (lower left panel) and the \textit{TESS} plus recent ground-based follow up observations (lower right panel). In these lower panels the continuous curves represent approximative analytic solution obtained with the formulae of \citet{borko15}. The residuals of the observed vs photodynamically modelled ETVs are plotted in the bottommost panels.} 
\label{fig:overall_etvs} 
\end{center}
\end{figure*}  % Figure 8

%%--------------------------------------------------------------
\section{Joint Physical and Dynamical modelling of all the available observational data}
\label{sec:dyn}

We carried out complex photodynamical modelling of TIC\,209409435 with the use of the software package {\sc Lightcurvefactory} \citep[see][and further references therein]{borko19a,borko20}. In order to obtain a comprehensive, as well as physically and dynamically consistent model for this triple we simultaneously analysed: (i) four sets of photometric lightcurves (the two sectors of \textit{TESS} measurements, the historical WASP observations, and T.\,G.~Tan's Cousins $R_\mathrm{C}$-band and F.-J.~Hambsch's Johnson $V$-band observations obtained during our photometric follow-up campaign); (ii) both the primary and secondary ETV curves deduced from these lightcurves (see Table\,\ref{Tab:TIC_209409435_ToM}); and also (iii) the archived stellar energy distribution (SED) for this system, in the form of different photometric passband magnitudes (see in Table\,\ref{tbl:mags}). 

The advantages of such a simultaneous, comprehensive analysis, as well as the consecutive steps of the complete procedure were explained in detail in a series of papers \citep{rappaport17,borko18,borko19a,borko19b,borko20} and we believe that it is not necessary to repeat them here. Therefore, here we discuss only those details that are specific to the current study.

Before carrying out the photodynamical analysis we further prepared the four sets of lightcurves as described in the following three paragraphs.

For the photodynamical analysis of the {\it TESS} data we processed the original \textit{TESS} full-frame images using a convolution-based differential photometric pipeline, based on the various tasks of the {\sc Fitsh} package \citep{2012MNRAS.421.1825P}. Then, from the raw lightcurve obtained  from this process we removed what are likely instrumental effects using the software package  {\sc W{\={o}}tan} \citep{hippkeetal19}. We also restricted the intervals of time within the lightcurves to be modelled to the orbital phase domain $\pm0.04$ orbital cycles around each eclipse.  The major exceptions to this were the regions around the third-body eclipses.  For these, we also utilized the flat, out-of-eclipse lightcurves between the two consecutive regular eclipses preceding and following the two pairs of third-body eclipses (see in Fig.\,\ref{fig:outecl}).

Regarding the WASP measurements, we utilized data from the entire time domain because even the out-of-eclipse regions for the inner binary also carried potentially significant information about where the third body eclipses might have occurred. However, in order to reduce computational costs we formed 1-hr averages from these out-of-eclipse data points, and these binned data were then used for the analysis.

Similarly, in the case of the ground-based photometric follow-up observations, instead of the original data, we used their 15-minute averages. %Furthermore, for the sake of homogeneity, we used only the Cousins $R_\mathrm{C}$-band observations of TG Tan. These data contain three fully covered and two partially observed regular, inner eclipses, as well as sections of two of the three outer eclipses that occurred in February and March 2020.

In the first stage of our study we carried out a joint, simultaneous photodynamical analysis of the prepared \textit{TESS} and WASP lightcurves, as well as the first few ground-based regular eclipse observations of T.\,G.~Tan, together with the ETV curves derived from them.  From this analysis, we were able to determine not only the outer period, together with the other orbital elements of the outer orbit, but also well constrained relative (i.e., dimensionless) stellar parameters (i.e., fractional radii and ratios of temperatures and masses).   With these parameters in hand we were thereby able to predict the occurrence times of the forthcoming extra eclipses with sufficient accuracy to guide us in making further ground-based follow-up observations.  

In the absence of radial velocity measurements, however, we were unable to determine model-independent, dynamical masses for each component.  Therefore, in the second stage of the analysis, in order to obtain physical quantities within the framework of a self-consistent model, similar to the method followed in \citet{borko20}, we added the observed composite SED of the triple system to the fit, and made efforts to find consistent, coeval \texttt{PARSEC} evolutionary tracks \citep{PARSEC} for the three stars. 

For this latter purpose we used machine readable \texttt{PARSEC} isochrone tables generated via the web based tool CMD 3.3\footnote{\url{http://stev.oapd.inaf.it/cgi-bin/cmd}}. These tables contain theoretically computed fundamental stellar parameters and absolute passband magnitudes in several different photometric systems, for a large three-dimensional grid of ages, metallicities and initial stellar masses \citep[see also][and further references therein]{chenetal19}. The preselection of the most appropriate subsets of the tabulated isochrones for the current analysis, as well as the interpolation of fundamental stellar parameters and absolute passband magnitudes from the grid of values for an actual trial run is described in Section 3 of \citet{borko20}. 

The \texttt{PARSEC} tracks during each trial step were used as follows. The actual trial values of stellar mass, age, and metallicity determined the position of each star on the set of \texttt{PARSEC} tracks. Then, using a trilinear interpolation based on the closest grid points of the pre-calculated tables, the code interpolated the radius and temperature of each star as well as their absolute passband magnitudes for the SED fitting. These stellar radii and temperatures were used as input parameters to the light curve modelling section of the code. Furthermore, the interpolated absolute passband magnitudes transformed into model passband magnitudes  with the use of the extinction parameter and the system's distance.  Then, their sum was compared to the observed magnitudes in each passband. Note, however, in the last steps, the distance ($d$) was not a free parameter, but was constrained a posteriori in each trial step by minimizing the value of $\chi^2_\mathrm{SED}$. We intentionally did not use the Gaia DR2 parallax for constraining the distance. The reason is that the Gaia DR2 data might contain systematics for compact triple systems \citep[see, e.g. the discussion of this problem in][]{borko20}. However, as we will discuss above, our results are in perfect agreement with the Gaia DR2 catalog data.

Therefore, in most of the runs we adjusted the following parameters:
\begin{itemize}
\item[--]{Three parameters related to the orbital elements of the inner binary as follows: the eccentricity and the argument of periastron via $e_1\cos\omega_1$ and $e_2\sin\omega_2$,  and the inclination, $i_1$.}  
\item[--]{Six parameters related to the orbital elements of the wide orbit of the third component: $P_2$, $e_2\sin\omega_2$, $e_2\cos\omega_2$, $i_2$, the time of the superior conjunction of the outer star, $\mathcal{T}^\mathrm{sup}_2$, and the position angle of the node of the wide orbit, $\Omega_2$. }\footnote{As $\Omega_1=0\degr$ was assumed at epoch $t_0$ for all runs, $\Omega_2$ set the initial trial value of the differences of the nodes (i.~e., $\Delta\Omega$), which is the truly relevant parameter for dynamical modelling.}
\item[--]{Three mass-related parameters: the mass of the primary of the inner binary, $m_\mathrm{A}$, and the mass ratios of the two orbits $q_{1,2}$.}
\item[--]{Four lightcurve-dependent parameters as the passband-dependent `third light(s)' $\ell_{TESS}$, $\ell_\mathrm{WASP}$, $\ell_{R_C}$ and $\ell_V$.}
\item[--]{Three parameters for the \texttt{PARSEC} isochrone and SED fitting: the age, $\log\tau$, and the metallicity, $[M/H]$, of the system and the extinction coefficient $E(B-V)$.} 
\end{itemize} 

Furthermore, the following parameters were internally constrained:  
\begin{itemize}
\item[--]{The instantaneous orbital period, $P_1$, of the inner binary and the inferior conjunction time, $\mathcal{T}^\mathrm{inf}_1$, of the secondary component of the inner binary, i.e., the mid-primary-eclipse-time at the zero epoch, were constrained via the use of the ETV curves in the manner explained in Appendix\,A of \citet{borko19a};} 
\item[--]{The stellar radii, $R_\mathrm{A,B,C}$, and effective temperatures, $T_\mathrm{A,B,C}$, were calculated from interpolation at each trial step from the appropriate grid elements of the \texttt{PARSEC} isochrone tables;}
\item[--]{and, the distance of the system was constrained a posteriori by minimizing the value of $\chi^2_\mathrm{SED}$.}
\end{itemize}

Regarding the other lightcurve-related parameters, we utilized a logarithmic limb-darkening law.  The limb-darkening coefficients were interpolated from passband-dependent tables in the {\tt Phoebe} software \citep{Phoebe}. In turn, the {\tt Phoebe} tables were derived from the \citet{castellikurucz04} stellar atmospheric models. We have found that due to the nearly spherical shapes of the stars in the inner binary, accurate settings of the gravity darkening coefficients have almost no effect on the lightcurve solution. Therefore, we simply adopted a fixed value of  $g = 0.32$ which, according to the venerable model of \citet{lucy67}, is appropriate for stars with a convective envelope. Furthermore, we neglected the reradiation/illumination effect in order to save computation time.  By contrast, the Doppler-boosting effect \citep{loebgaudi03,vankerkwijketal10}, which is also found to be negligible for this system, requires only very minor additional computational time, and was therefore included in the model calculations.
 
\begin{figure*}
\begin{center}
\includegraphics[width=0.47 \textwidth]{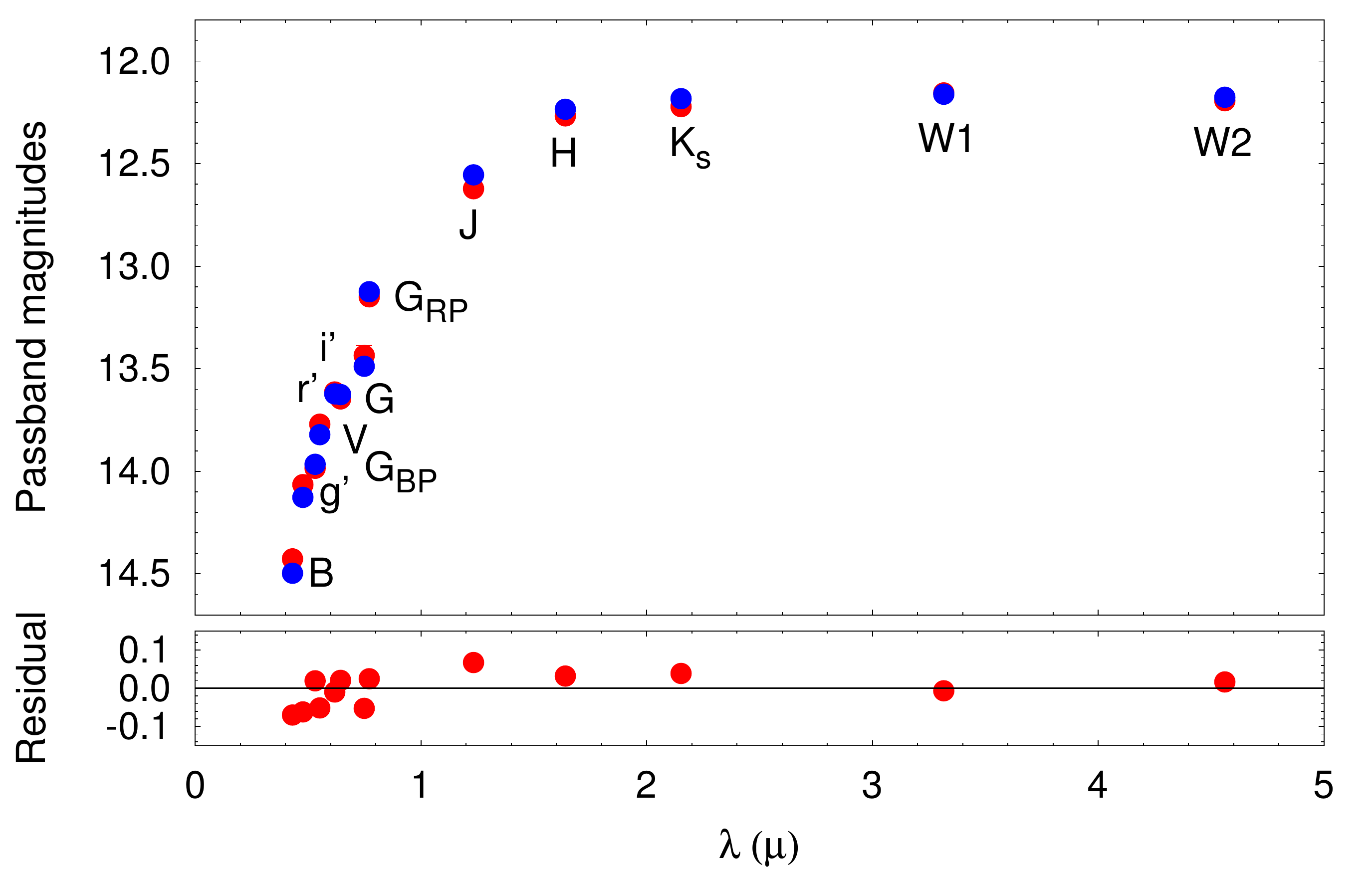}  
\includegraphics[width=0.49 \textwidth]{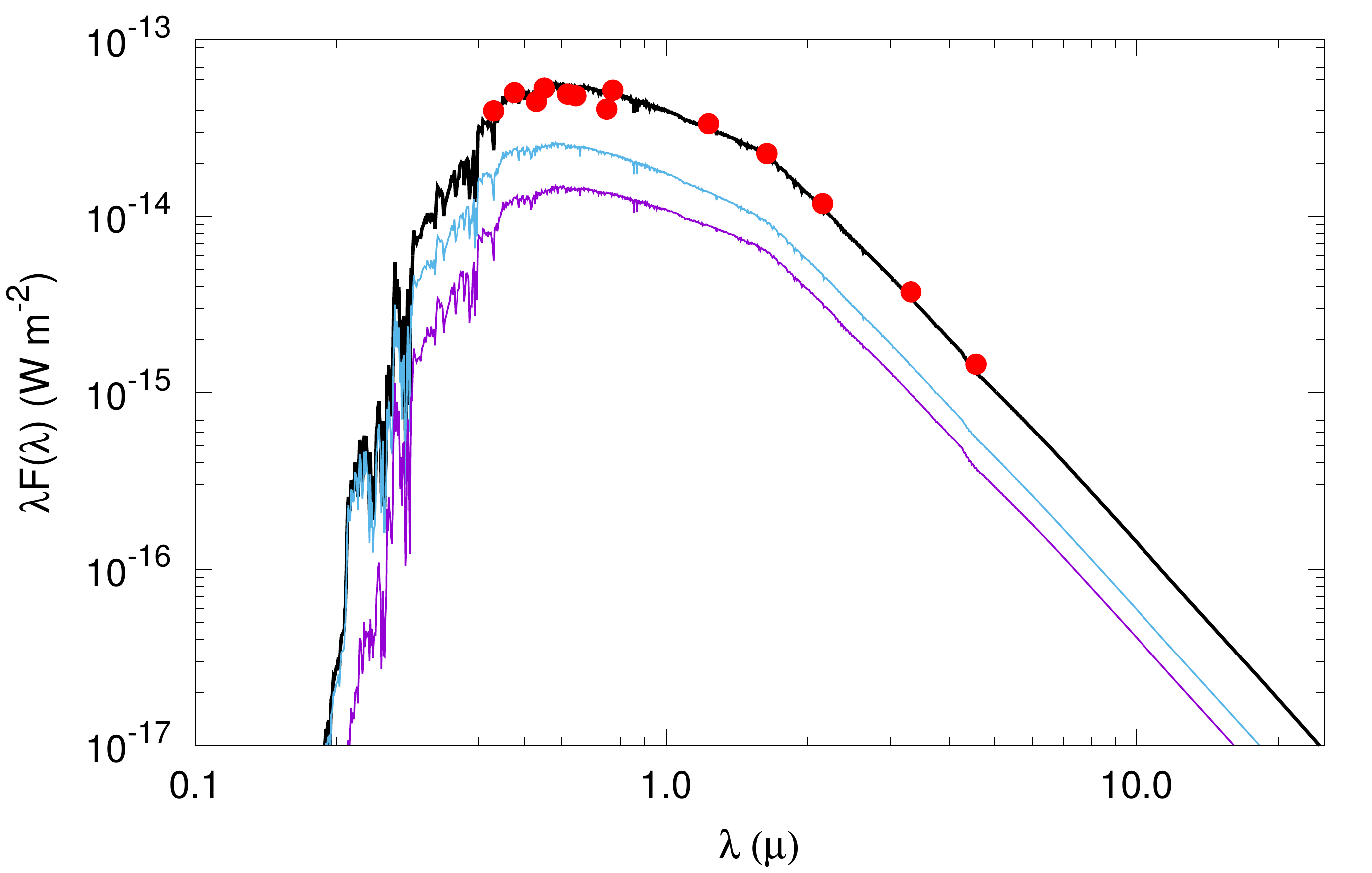}
\caption{The summed SED of the three stars of TIC\,209409435 both in the magnitude and the flux domains. The {\it left panel} displays the cataloged values of the passband magnitudes (red filled circles; tabulated in Table\,\ref{tbl:mags}) versus the model passband magnitudes derived from the absolute passband magnitudes interpolated with the use of the \texttt{PARSEC} tables (blue filled circles).  In the {\it right panel} the dereddened observed magnitudes are converted into the flux domain (red filled circles), and overplotted with the quasi-continuous summed SED for the triple star system (thick black line). This SED is computed from the \citet{castellikurucz04} ATLAS9 stellar atmospheres models (\url{http://wwwuser.oats.inaf.it/castelli/grids/gridp00k2odfnew/fp00k2tab.html}). The separate SEDs of the more massive third component and the twin stars of the inner binary are also shown with thin green and purple lines, respectively. } 
\label{fig:sedfit} 
\end{center}
\end{figure*}

The orbital and astrophysical parameters derived from the photodynamical analysis are tabulated in Table\,\ref{tab: syntheticfit_TIC209} and will be discussed in the subsequent Section\,\ref{sec:par}. The corresponding model lightcurves are presented in Figs.\,\ref{fig:outecl}, \ref{fig:wasp_lc}, and \ref{fig:lc_ground}, while the model ETV curves plotted against the observed ETVs are shown in Fig.\,\ref{fig:overall_etvs}. Finally, the results of the  stellar SED are plotted in Fig.\,\ref{fig:sedfit}.

\begin{table*}
 \centering
\caption{Orbital and astrophysical parameters of TIC\,209409435 from the joint photodynamical lightcurve, ETV, SED and \texttt{PARSEC} isochrone solution. Besides the usual observational system of reference related angular orbital elements ($\omega$, $i$, $\Omega$), their counterparts in the system's invariable plane related dynamical frame of reference are also given ($\omega^\mathrm{dyn}$, $i^\mathrm{dyn}$, $\Omega^\mathrm{dyn}$). Moreover, $i_\mathrm{m}$ denotes the mutual inclination of the two orbital planes, while $i_\mathrm{inv}$ and $\Omega_\mathrm{inv}$ give the position of the invariable plane with respect to the tangential plane of the sky (i.\,e., in the observational frame of reference). }
 \label{tab: syntheticfit_TIC209}
\begin{tabular}{@{}llll}
\hline
\multicolumn{4}{c}{orbital elements$^a$} \\
\hline
   & \multicolumn{3}{c}{subsystem}  \\
   & \multicolumn{2}{c}{A--B} & AB--C  \\
  \hline
  $P$ [days] & \multicolumn{2}{c}{$5.717471_{-0.000021}^{+0.000027}$} & $121.8723_{-0.0009}^{+0.0010}$   \\
  $a$ [R$_\odot$] & \multicolumn{2}{c}{$16.35_{-0.12}^{+0.29}$} & $145.3_{-1.1}^{+2.7}$ \\
  $e$ & \multicolumn{2}{c}{$0.00407_{-0.00005}^{+0.00005}$} & $0.39653_{-0.00012}^{+0.00013}$ \\
  $\omega$ [deg]& \multicolumn{2}{c}{$154.5_{-3.4}^{+3.9}$} & $195.32_{-0.07}^{+0.08}$ \\ 
  $i$ [deg] & \multicolumn{2}{c}{$89.978_{-0.072}^{+0.089}$} & $89.795_{-0.011}^{+0.013}$ \\
  $\tau$ [BJD - 2400000]& \multicolumn{2}{c}{$583386.976_{-0.054}^{+0.063}$} & $58295.950_{-0.019}^{+0.021}$ \\
  $\Omega$ [deg] & \multicolumn{2}{c}{$0.0$} & $0.113_{-0.126}^{+0.109}$ \\
  $i_\mathrm{m}$ [deg] & \multicolumn{3}{c}{$0.243_{-0.080}^{+0.082}$} \\
  $\omega^\mathrm{dyn}$ [deg]& \multicolumn{2}{c}{$187_{-37}^{+25}$} & $54_{-32}^{+26}$ \\
  $i^\mathrm{dyn}$ [deg] & \multicolumn{2}{c}{$0.201_{-0.067}^{+0.068}$} & $0.042_{-0.014}^{+0.014}$\\
  $\Omega^\mathrm{dyn}$ [deg] & \multicolumn{2}{c}{$148_{-23}^{+36}$} & $328_{-23}^{+36}$ \\
  $i_\mathrm{inv}$ [deg] & \multicolumn{3}{c}{$89.826_{-0.013}^{+0.018}$} \\
  $\Omega_\mathrm{inv}$ [deg] & \multicolumn{3}{c}{$0.094_{-0.104}^{+0.090}$} \\
  \hline
  mass ratio $[q=m_\mathrm{sec}/m_\mathrm{pri}]$ & \multicolumn{2}{c}{$1.002_{-0.003}^{+0.003}$} & $0.546_{-0.004}^{+0.003}$ \\
  $K_\mathrm{pri}$ [km\,s$^{-1}$] & \multicolumn{2}{c}{$72.46_{-0.60}^{+1.34}$} & $23.20_{-0.23}^{+0.48}$  \\ 
  $K_\mathrm{sec}$ [km\,s$^{-1}$] & \multicolumn{2}{c}{$72.29_{-0.53}^{+1.23}$} & $42.55_{-0.70}^{+1.16}$ \\ 
  \hline  
\multicolumn{4}{c}{stellar parameters} \\
\hline
   & A & B &  C  \\
  \hline
 \multicolumn{4}{c}{Relative quantities$^b$} \\
  \hline
 fractional radius [$R/a$]  & $0.0533_{-0.0004}^{+0.0004}$ & $0.0535_{-0.0004}^{+0.0004}$  & $0.00705_{-0.00007}^{+0.00007}$ \\
 fractional flux [in \textit{TESS}-band] & $0.2616$  & $0.2654$    & $0.4307$ \\
 fractional flux [in WASP-band]& $0.2613$  & $0.2657$    & $0.4566$ \\
 fractional flux [in $R_\mathrm{C}$-band]& $0.2667$  & $0.2708$    & $0.4484$ \\
 fractional flux [in $V$-band]& $0.2648$  & $0.2692$    & $0.4626$ \\
 \hline
 \multicolumn{4}{c}{Physical Quantities} \\
  \hline 
 $m$ [M$_\odot$] & $0.895_{-0.019}^{+0.048}$ & $0.897_{-0.020}^{+0.050}$ & $0.976_{-0.023}^{+0.058}$ \\
 $R^b$ [R$_\odot$] & $0.872_{-0.011}^{+0.020}$ & $0.875_{-0.013}^{+0.021}$ & $1.027_{-0.015}^{+0.015}$ \\
 $T_\mathrm{eff}^b$ [K]& $5769_{-96}^{+74}$ & $5779_{-101}^{+73}$ & $6074_{-87}^{+80}$ \\
 $L_\mathrm{bol}^b$ [L$_\odot$] & $0.753_{-0.053}^{+0.076}$ & $0.762_{-0.056}^{+0.081}$ & $1.277_{-0.068}^{+0.134}$ \\
 $M_\mathrm{bol}^b$ & $5.08_{-0.10}^{+0.08}$ & $5.07_{-0.11}^{+0.08}$ & $4.50_{-0.11}^{+0.06}$ \\
 $M_V^b           $ & $5.15_{-0.12}^{+0.09}$ & $5.13_{-0.12}^{+0.09}$ & $4.54_{-0.12}^{+0.06}$ \\
 $\log g^b$ [dex] & $4.509_{-0.005}^{+0.005}$ & $4.507_{-0.005}^{+0.005}$ & $4.406_{-0.013}^{+0.013}$ \\
 \hline
$\log$(age) [dex] &\multicolumn{3}{c}{$9.672_{-0.250}^{+0.096}$} \\
$[M/H]$  [dex]    &\multicolumn{3}{c}{$-0.155_{-0.236}^{+0.054}$} \\
$E(B-V)$ [mag]    &\multicolumn{3}{c}{$0.048_{-0.029}^{+0.013}$} \\
extra light $\ell_4$  [in \textit{TESS}-band]&\multicolumn{3}{c}{$0.084_{-0.013}^{+0.011}$} \\
extra light $\ell_4$  [in WASP-band]&\multicolumn{3}{c}{$0.034_{-0.021}^{+0.037}$} \\
extra light $\ell_4$  [in $R_\mathrm{C}$-band]&\multicolumn{3}{c}{$0.008_{-0.006}^{+0.010}$} \\
extra light $\ell_4$  [in $V$-band]&\multicolumn{3}{c}{$0.030_{-0.014}^{+0.015}$} \\
$(M_V)_\mathrm{tot}^b$           &\multicolumn{3}{c}{$3.71_{-0.12}^{+0.08}$} \\
distance [pc]                &\multicolumn{3}{c}{$990_{-17}^{+22}$}  \\  
\hline
\end{tabular}

\textit{Notes. }{$a$: Instantaneous, osculating orbital elements, calculated for epoch $t_0=2458382.0000$ (BJD); $b$: Interpolated from the \texttt{PARSEC} isochrones.}
\end{table*}

  % Figure 9

% Add somewhere:  http://wwwuser.oats.inaf.it/castelli/grids/gridp00k2odfnew/fp00k2tab.html

%%--------------------------------------------------------------
\section{Discussion of the results}
\label{sec:par}

According to our results, TIC\,209409435 is comprised of three very similar solar type stars.  The two components of the inner binary pair are perfect twins, having a mass ratio of $q_1=1.002\pm0.003$, while the outer star has a larger mass only $\sim$9\% higher than the inner components. These mass ratios are very robust and model independent due to the photodynamical part of our analysis. By contrast, in the absence of radial velocity data, the actual physical masses of the stars (and, correspondingly, other fundamental parameters such as the physical dimensions of the components), were obtained in an astrophysical model-dependent way, with the use of theoretical stellar evolutionary tracks in the form of \texttt{PARSEC} isochrones.  Our combined analysis, however, has led to a self-consistent and robust result whose high fidelity is confirmed by the comparison of the observed net stellar SED to the model SED. This  yields a system distance of $d=990\pm20$\,pc, which is about to 2-$\sigma$ from the Gaia DR2 distance of $d_\mathrm{GaiaDR2}=949\pm15$\,pc. Note, however, that the triple nature of this source might have resulted in systematic discrepancies in the Gaia DR2 parallax \citep[see, e.~g.][]{benedictetal18}. 

From this same analysis we can state with high confidence that TIC\,209490435 is a system having three MS stars with masses $m_\mathrm{A,B}=0.90\pm0.03\,\mathrm{M}_\odot$ and $m_\mathrm{C}=0.98\pm0.04\,\mathrm{M}_\odot$, with effective temperatures $T_\mathrm{A,B}=5800\pm100$\,K and $T_\mathrm{C}=6070\pm90$\,K.  The system is $\sim5_{-2}^{+1}$\,Gyr-old, and slightly metal-deficient with $[M/H]=-0.16\pm0.1$. 

We have found that the triple is very flat, where the mutual inclination of the inner and outer orbits is $i_\mathrm{m}=0\fdg24\pm0\fdg08$. The flatness of the system together with the almost circular inner, and moderately eccentric outer orbit, should make the configuration of this triple stable over the nuclear lifetime of the stars. We have taken a step toward verifying this with a 10\,Myr-long numerical integration (carried out with the same integrator that was used for the photodynamical modelling). From an observational perspective, in the near complete absence of orbital plane precession, the inner pair will remain an eclipsing binary, and the outer star will also continue to produce the extra third-body eclipses over any human time scales. During the 10\,Myr-long integration, the inner and outer orbital inclinations oscillated between $89\fdg64\lesssim i_1\lesssim90\fdg01$, and $89\fdg79\lesssim i_2\lesssim89\fdg86$ with a period of $P_\mathrm{prec}\approx17$\,yr.\footnote{One should keep in mind, however, that in the absence of any information on the axial rotation of the stars in the binary, we assumed synchronized and aligned rotation for the integration. Inclined spin axes might result in some orbital plane precession, and even drastic variations in the configuration of the triple system on a longer time scale \citep[see, e.g.][]{correiaetal16}.} {The apsidal precession of the low-eccentricity inner binary orbit has a very similar period of $P_\mathrm{apse1}\approx15$\,yr, while the outer orbit has a century-long $P_\mathrm{apse2}\approx103$\,yr apsidal period. The short-term variations in some of the orbital elements of the inner and outer orbits, obtained from our numerical integration, are plotted in Fig.\,\ref{fig:orbelements_numint}.

\begin{figure*}
\begin{center}
\includegraphics[width=0.32 \textwidth]{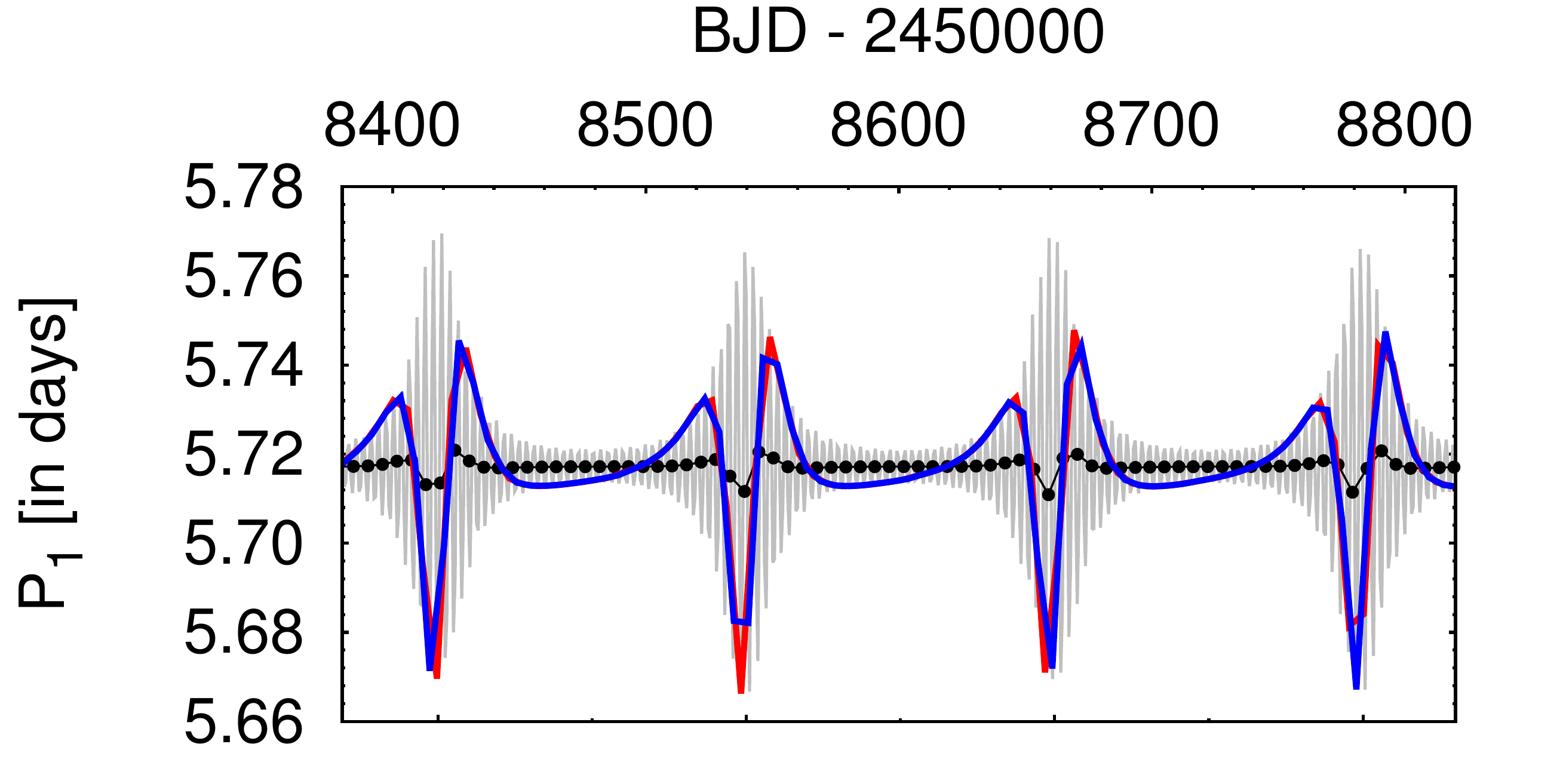}\includegraphics[width=0.32 \textwidth]{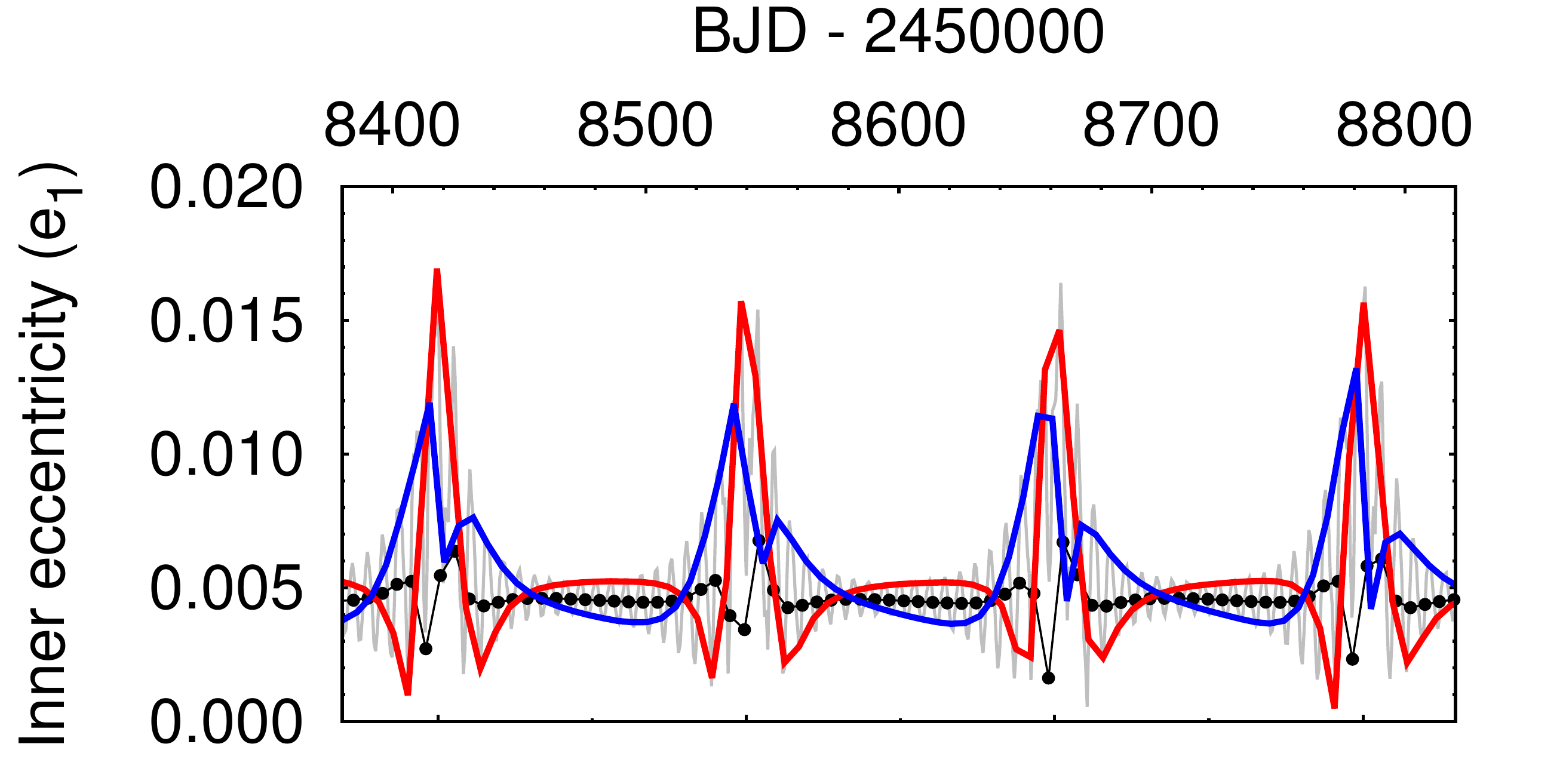}\includegraphics[width=0.32 \textwidth]{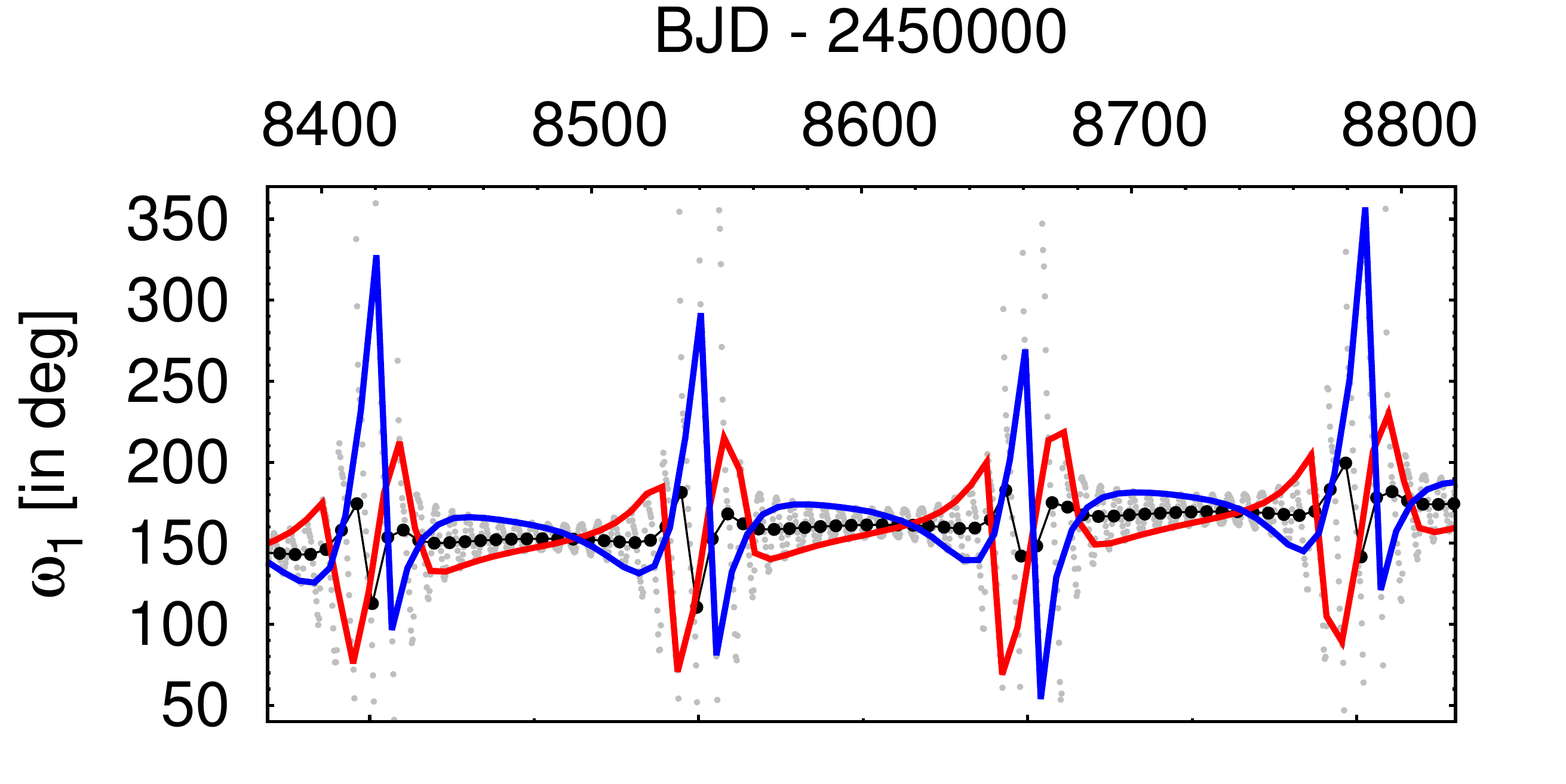}
\includegraphics[width=0.32 \textwidth]{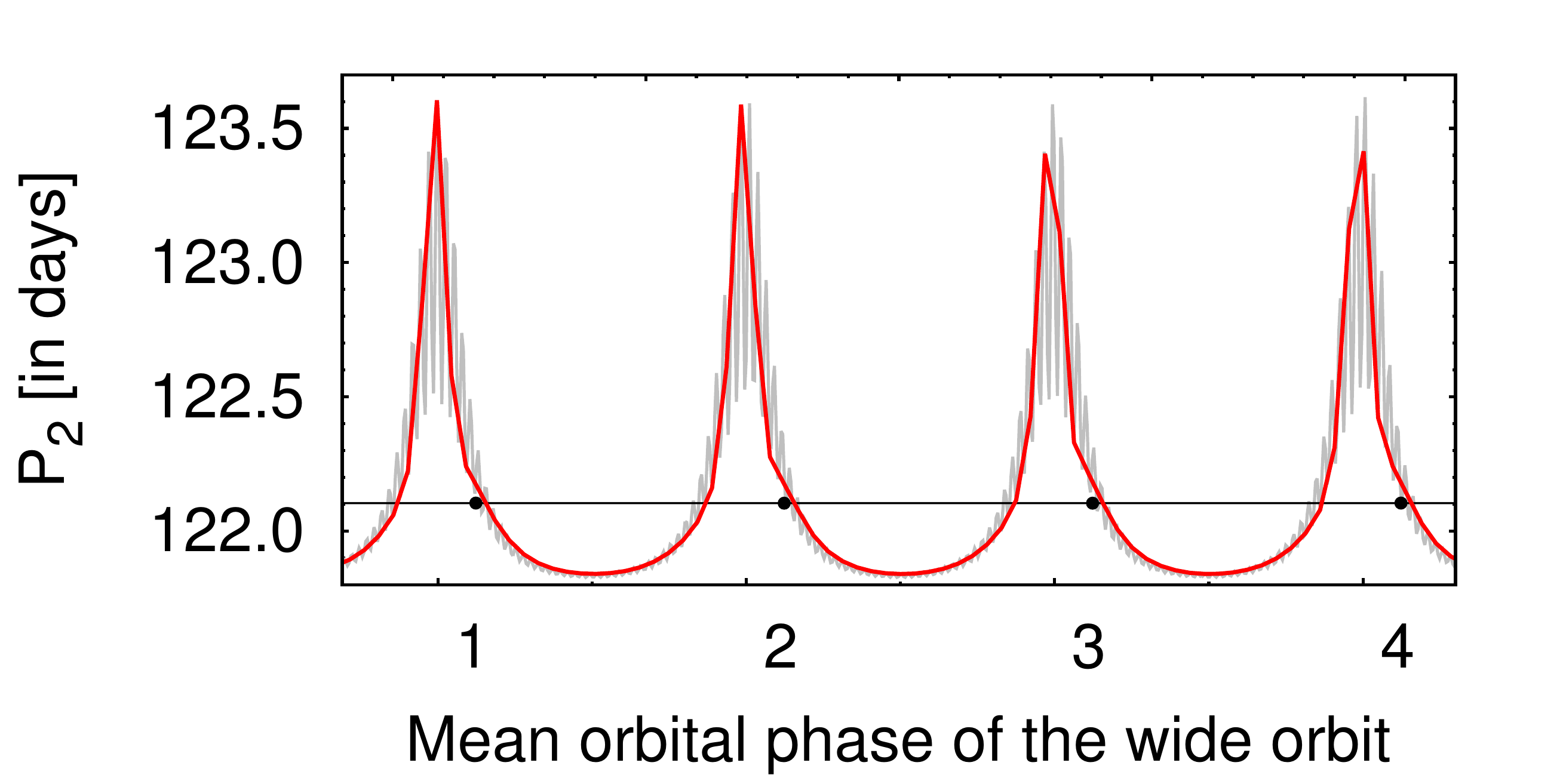}\includegraphics[width=0.32 \textwidth]{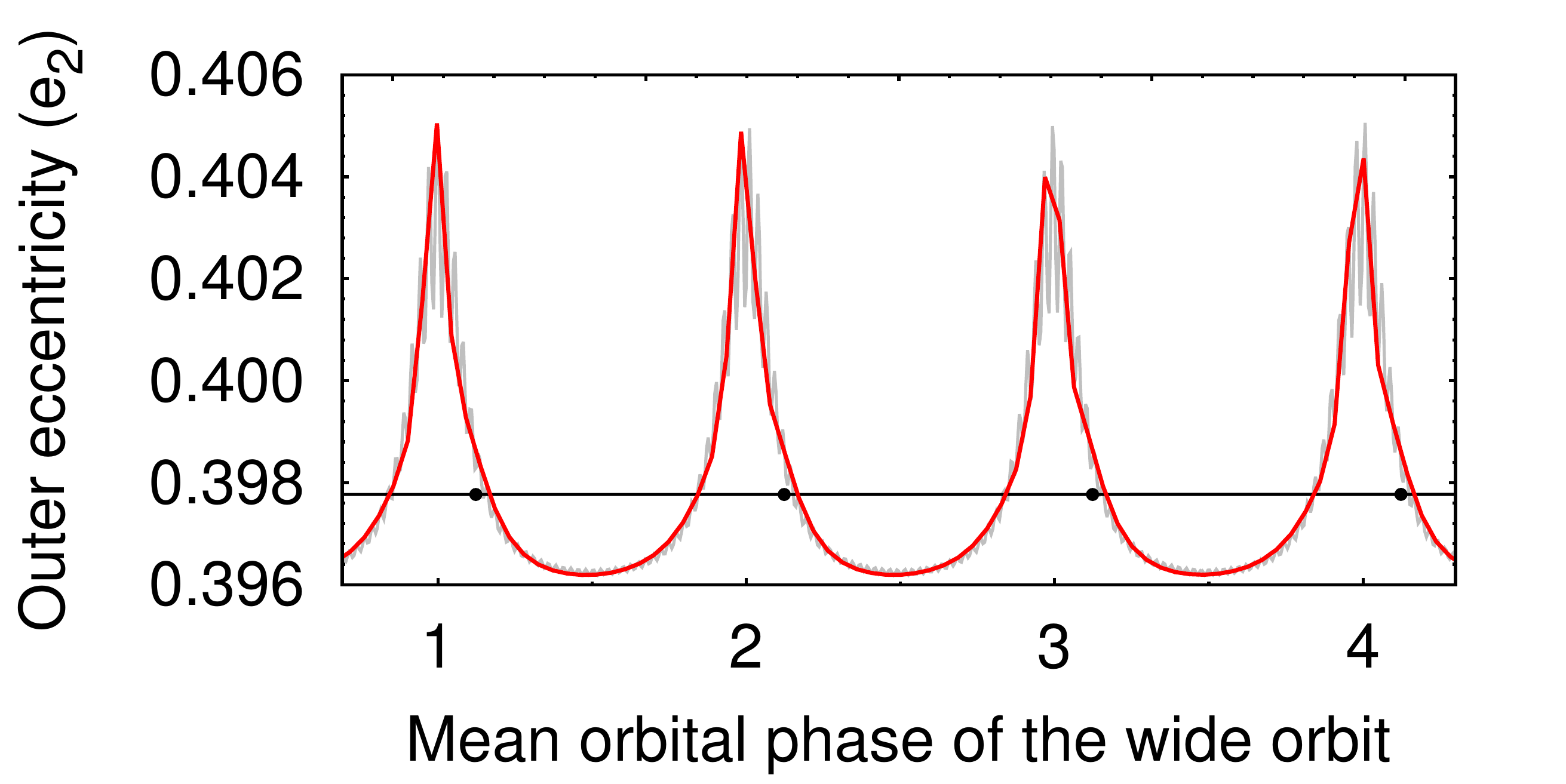}\includegraphics[width=0.32 \textwidth]{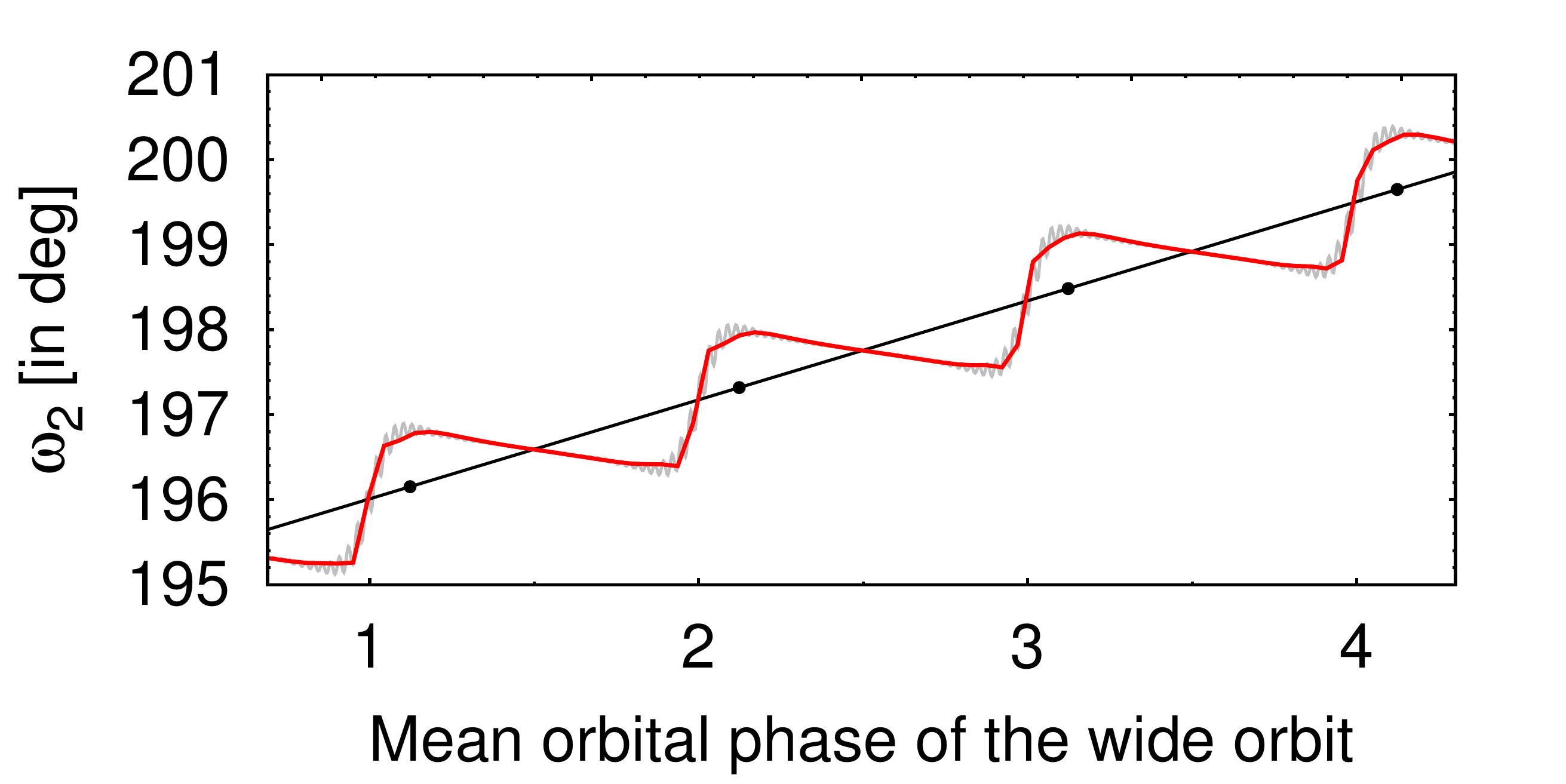}
\caption{Variations of the instantaneous (osculating) anomalistic periods, eccentricities and (observable) arguments of periastron of the inner and outer binaries (grey curves in all panels), obtained via numerical integration, together with the one-inner/outer-orbit averages (black).  Furthermore, the red and blue lines connect the instantaneous orbital elements sampled at those integration points that are closest to the primary and secondary mid-eclipse points of the inner binary, respectively. (Note, for the outer orbit the red and blue lines coincide almost perfectly, therefore, we plot only the red one.)}
\label{fig:orbelements_numint} 
\end{center}
\end{figure*}  % Figure 10

Turning back to the most prominent characteristic of our triple, i.e. its flatness,} other similarly very flat ($i_\mathrm{m}<5\degr$) and compact ($P_2<200$\,d) triple systems with accurately known parameters that were reported previously are HD\,181068 \citep{borko13}, HD\,144548 \citep{alonsoetal15}, $\xi$\,Tau \citep{nemravovaetal16}, EPIC\,249432662 \citep{borko19a}, HIP\,41431 \citep{borko19b} and TIC\,220397947 \citep{borko20}.\footnote{Note, \citet{borko16} lists 8 additional compact triples with $P_2<200$ days in the original \textit{Kepler}-field with mutual inclinations most probably less than $5\degr$.  But, in the absence of accurate photodynamical solutions, we do not list these systems. Similarly, we omit $\lambda$ Tau, the triple with the shortest outer period known \citep{ebbighausenstruve56,fekeltomkin82}. Though, in the absence of any eclipse depth variations over the last 110 years one can suppose with a great certainty that this triple is also extremely flat \citep[see, e.~g.][]{soderhjelm75,kiselevaetal98,berdyuginetal18}. Finally we note that we made a concerted effort to collect accurate photodynamical solutions from the literature for similarly flat systems with somewhat longer outer periods of up to $P_2\leq1000$ days, but still relatively compact.  Unfortunately, we were unsuccessful in finding any.} %this system also needs a further accurate photodynamical modelling.}.  
All of these systems show remarkable similarities, as one can see in Table\,\ref{tab:comp}.  For example, all systems have high mass-ratio inner binaries ($q_1\geq0.88$), and in five of the seven the inner pair is formed by near twin stars (i.e. $q_1> 0.95$).  Furthermore, in all but one of the triples the outer third component is the more massive. Moreover, all inner orbits are almost circular, while the outer orbits, apart from HD\,181068, have moderate eccentricities.\footnote{In this latter triple the outer component is a red giant, therefore, the circularization of the outer orbit can probably be explained with the more effective tidal damping effects in its present red giant state.} Such flat triple systems were most likely formed by disk fragmentation and accretion-driven migration.  However, the quantitative details of these effects have not been fully explored \citep[see, e.~g.][]{tokovininmoe20}, and it is also unclear what the role is of the third star in this process. Therefore, it is especially important to improve the sample of compact flat triple systems with accurately known system parameters.

\begin{table*}
 \centering
\caption{Comparison of parameters of flat, compact hierarchical triple stars}
 \label{tab:comp}
\begin{tabular}{@{}llllllllll}
\hline
  & $P_1$ & $P_2$ & $e_1$ & $e_2$ & $i_\mathrm{m}$ & $q_1$ & $q_2$ & $m_\mathrm{A}$ &  $m_\mathrm{C}/m_\mathrm{A}$\\
\hline
HD\,144548     & 1.63 & 33.95 &$\lesssim0.0015$& 0.265 & 0.2 & 0.96 & 0.75 & 0.98 & 1.47 \\
HD\,181068     & 0.91 & 45.47 & 0.0            & 0.0   & 0.8 & 0.95 & 1.68 & 0.92 & 3.28 \\
HIP\,41431     & 2.93 & 59.16 & 0.0087         & 0.278 & 2.16& 0.98 & 0.51 & 0.63 & 1.01 \\
TIC\,220397947 & 3.55 & 77.08 & 0.0011         & 0.225 & 0.57& 0.95 & 0.25 & 1.15 & 0.49 \\
TIC\,209409435 & 5.72 & 121.9 & 0.0041         & 0.397 & 0.24& 1.00 & 0.55 & 0.90 & 1.09 \\ 
$\xi$\,Tau     & 7.15 & 145.2 & 0.0000         & 0.197 & 2.50& 0.90 & 0.88 & 2.23 & 1.67 \\
EPIC\,249432662& 8.19 & 188.4 & 0.0034         & 0.221 & 0.17& 0.88 & 1.14 & 0.45 & 2.14 \\ 
\hline   
\end{tabular}

\textit{Note. } HIP\,41431, $\xi$\,Tau and (probably) TIC\,220397947 have further, more distant stellar components, i.e., these are (at least) 3+1 hierarchical quadruple systems.
\end{table*}

\section{Summary and Conclusions}
\label{sec:disc}

In this work we report on the discovery and analysis of a new triply eclipsing triple star system, TIC 209409435.  The discovery observations were made with \textit{TESS} during Sectors 3 and 4.  We also made use of archival WASP data as well as follow-up ground-based observations by amateur astronomers.

We carried out a photodynamical analysis of the observational data which included the lightcurves, ETVs extracted from the lightcurves, the SEDs, and \texttt{PARSEC} evolution tracks. From this analysis we were able to obtain a comprehensive set of system masses and orbital parameters, all without being able to make RV observations.

In spite of the hundreds of thousands of eclipsing binaries that have been discovered over the years, including many whose light curves  have been very well studied with CoRoT, \textit{Kepler}, \textit{K2}, and \textit{TESS}, there are only 17 known EBs that have eclipsing third bodies orbiting them, including this new discovery (see Table\,\ref{tbl:tripleeclipsers}). TIC 209409435, among the 17 triply eclipsing triples, has the 6th shortest outer orbital period, the 4th highest ratio of $P_1/P_2$, and the 6th highest ratio of $P_1^2/P_2$, the latter being a measure of the dynamical time delays that are produced.

Thus TIC 209409435 is an impressive interactive system with pronounced ETVs and exotic-looking third-body eclipses twice every 122 days.  The presence of the outer body eclipses helped enable us to make very robust determinations of the orbital parameters and the masses (see  Table \ref{tab: syntheticfit_TIC209}).  The stellar masses and radii are good to about 3\% accuracy on average.  The triple system is extremely flat and aligned with the observers with $i_1 =  89.98^\circ \pm 0.08^\circ$, $i_2 =  89.79^\circ \pm 0.09^\circ$, and mutual inclination $i_{\rm m} =  0.24^\circ \pm 0.08^\circ$.  The system is 5 Gyr old and is manifestly highly dynamically stable.  The photometric distance of $990 \pm 20$ pc matches the Gaia distance of $949 \pm 15$ pc within 2-$\sigma$.

The photodynamical analysis has led to an orbital ephemeris that should be able to accurately predict when future third-body events can be observed from the ground, including with small telescopes.  Moreover, TIC\,209409435 is scheduled to be observed again in \textit{TESS} Sectors 30 and 31 (i.e., nominally, between 22 Sept and 19 Nov 2020) during the extended Year 3 \textit{TESS} mission. During this interval, extra third-body eclipsing events are expected on 1--3 October and 3--4 November. Though these dates are somewhat inauspicious in the sense that they are close to the mid-times of the two sectors (i.e. to the data downloading time), some portions of these events will hopefully be observed, allowing us to further refine the photodynamical model.

Overall we have found that TIC 209409435 has been a very gratifying system to have worked on.

\section*{Acknowledgments}

T.\,B. and T.\,M. acknowledge the financial support of the Hungarian National Research, Development and Innovation Office -- NKFIH Grant KH-130372.

This paper includes data collected by the \textit{TESS} mission. Funding for the \textit{TESS} mission is provided by the NASA Science Mission directorate. Some of the data presented in this paper were obtained from the Mikulski Archive for Space Telescopes (MAST). STScI is operated by the Association of Universities for Research in Astronomy, Inc., under NASA contract NAS5-26555. Support for MAST for non-HST data is provided by the NASA Office of Space Science via grant NNX09AF08G and by other grants and contracts.

This work has made use  of data  from the European  Space Agency (ESA)  mission {\it Gaia}\footnote{\url{https://www.cosmos.esa.int/gaia}},  processed  by  the {\it   Gaia}   Data   Processing   and  Analysis   Consortium   (DPAC)\footnote{\url{https://www.cosmos.esa.int/web/gaia/dpac/consortium}}.  Funding for the DPAC  has been provided  by national  institutions, in  particular the institutions participating in the {\it Gaia} Multilateral Agreement.

This publication makes use of data products from the Wide-field Infrared Survey Explorer, which is a joint project of the University of California, Los Angeles, and the Jet Propulsion Laboratory/California Institute of Technology, funded by the National Aeronautics and Space Administration. 

This publication makes use of data products from the Two Micron All Sky Survey, which is a joint project of the University of Massachusetts and the Infrared Processing and Analysis Center/California Institute of Technology, funded by the National Aeronautics and Space Administration and the National Science Foundation.

We  used the  Simbad  service  operated by  the  Centre des  Donn\'ees Stellaires (Strasbourg,  France) and the ESO  Science Archive Facility services (data  obtained under request number 396301).

\end{document}